\documentclass[twocolumn]{aastex631}

\usepackage{amsmath} 
\usepackage{graphicx} 
\usepackage{textcomp}
\usepackage{gensymb}
\usepackage{natbib} 
\usepackage{longtable}
\usepackage{multirow}
\usepackage{float}
\usepackage{enumitem}

\begin{document}
\def\teff{$T_{eff}$}
\def\cs{$\chi^{2}$}
\def\rsun{$R_{\odot}$}
\def\msun{$M_{\odot}$}
\def\rstar{$R_{\star}$}
\def\rearth{$R_{\earth}$}
\def\av{$A_{V}$}
\def\kep{\textit{Kepler}}
\def\emcee{\texttt{emcee}}
\newcommand\val[3]{$#1^{+#2}_{-#3}$}

\graphicspath{{figures/}}
\shorttitle{Binary Planet Hosts I.}
\shortauthors{Sullivan, Kraus, \& Mann}

\title{Revising Properties of Planet-host Binary Systems I.: Methods and Pilot Study}

\author[0000-0001-6873-8501]{Kendall Sullivan}
\altaffiliation{NSF Graduate Research Fellow}
\affil{Department of Astronomy, University of Texas at Austin, Austin, TX 78712, USA}

\author{Adam L. Kraus}
\affil{Department of Astronomy, University of Texas at Austin, Austin, TX 78712, USA}

\author[0000-0003-3654-1602]{Andrew W. Mann}
\affiliation{Department of Physics and Astronomy, The University of North Carolina at Chapel Hill, Chapel Hill, NC 27599, USA}

\correspondingauthor{Kendall Sullivan}
\email{kendallsullivan@utexas.edu}

\begin{abstract}
To fully leverage the statistical strength of the large number of planets found by projects such as the \kep\ survey, the properties of planets and their host stars must be measured as accurately as possible. One key population for planet demographic studies is circumstellar planets in close binaries ($\rho < 50$au), where the complex dynamical environment of the binary inhibits most planet formation, but some planets nonetheless survive. Accurately characterizing the stars and planets in these complex systems is a key factor in better understanding the formation and survival of planets in binaries. Toward that goal, we have developed a new Markov Chain Monte Carlo fitting algorithm to retrieve the properties of binary systems using unresolved spectra, unresolved photometry, and resolved contrasts. We have analyzed 8 \kep\ Objects of Interest in M star binary systems using literature data, and have found that the temperatures of the primary stars (and presumed planet hosts) are revised upward by an average of 200 K. The planetary radii should be revised upward by an average of 20\% if the primary star is the host, and 80\% if the secondary star is the planet host. The average contrast between stellar components in the \kep\ band is 0.75 mag, which is small enough that neither star in any of the binaries can be conclusively ruled out as a potential planet host. Our results emphasize the importance of accounting for multiplicity when measuring stellar parameters, especially in the context of exoplanet characterization.
\end{abstract}


\section{Introduction}\label{sec:intro}
The \kep\ mission \citep{Borucki2010} was the one of the first space missions dedicated to detecting transiting exoplanets, and discovered more than 2,000 confirmed planets and another $\sim$4,000 planet candidates over its four-year \kep-prime mission. \kep\ was not a blind survey of the sky, but instead relied on the Kepler Input Catalog (KIC; \citealt{Brown2011}) and a complex set of magnitude cuts and other criteria to select which sources within the field of view would have data transferred back to Earth from the spacecraft \citep{Batalha2010}.

Although the KIC selection criteria excluded some binary stars from the target list, many \kep\ targets have been found to be multiples after the fact \citep[e.g.,][]{Law2014ApJ, Wang2014a, Kraus2016, Furlan2017}. The abundance of exoplanets coupled with the ubiquity of binaries ($\sim$50\% multiplicity rate for solar-type stars \citep{Raghavan2010}, decreasing to $\sim$25\% for M dwarfs \citep{Ward-Duong2015, Winters2019}), means that many known planets exist in binary systems. To fully leverage the wealth of data from the initial observations and subsequent follow-up and characterization of the more than 6,000 confirmed and candidate \kep\ planetary systems, the effects of multiplicity on measured planet parameters must be considered and corrected for.

The physical effects of multiplicity on planets are most important for close binary stars, but binaries affect observed planetary properties across a wide range of separations. Binaries on solar-system scales (projected separation $\rho \lesssim$ 50 au) suppress planet occurrence by at least two-thirds \citep{Wang2014a, Wang2014b, Kraus2016, Howell2021, Ziegler2021}), but some close binaries still successfully form and retain circumstellar planets \citep[planets on S-type orbits; e.g.,][]{Kraus2016}. Accurate stellar properties for binaries in this separation regime are vital for understanding both the suppression of planets at solar-system scales and the properties of the planets that survive in the hostile birth environment of a close binary.

At wider separations, stellar neighbors (either binaries or chance alignments) still affect inferred stellar (and planetary) parameters, even if the planet occurrence rate is not as sharply reduced. Intermediate-separation systems ($\rho \lesssim 2-3 \arcsec$) were unresolved in the KIC, so their measured stellar parameters were biased by the presence of the secondary. At even wider separations, stars with neighbors that fall inside the \kep\ pixels extracted to create a light curve ($\rho \lesssim 8-10 \arcsec$) are known to dilute the planetary signal by increasing the total stellar flux \citep{Ciardi2015}, meaning that the true size of the transiting object is larger than the size inferred from the transit depth. Dilution is typically corrected for using a flux ratio to create a factor that is then applied as a correction to the light curve and thus the apparent relative planet radius (R$_{p}$/\rstar). However, to derive an absolute planetary radius, the stellar radius must be known.

Beyond the well-understood effect of dilution, binaries also impact measured planetary properties in more subtle ways. For example, the increased system flux of a binary changes the inferred stellar luminosity and can bias the measured distance in systems where the parallax is poorly constrained. Contamination from a secondary star can change the measured effective temperature of a stellar spectrum \citep{El-Badry2018,Furlan2020, Sullivan2021}, changing estimates of the planet's irradiation and assessment of the planet's presence within the habitable zone of the star (HZ; the period-instellation regime in which liquid water would be possible on the planet surface; \citealt{Kasting1993}). Stellar radii are often estimated from models using the stellar temperature, so a modified \teff\ will also change the estimated stellar radius. Even if dilution of the light curve is accounted for when measuring a planet's parameters, an altered stellar radius will still change the inferred planet radius, which has implications for assessment of habitability and the presence (or lack thereof) of planets in the radius valley (a dearth of planets around 1.8 Earth radii; e.g., \citealt{Owen2013, Fulton2017, Petigura2022}). Additionally, a change in the stellar temperature will modify the inferred radiative environment of the planet. In an era of increasing interest in planetary demographics and progressively more accurate measurements, even small changes to planetary properties can be important for their inclusion in various data sets for calculating planetary statistics. 

Some individual planet-hosting binary systems have been characterized in detail, including measurement of the individual stellar parameters and assessment of the effect of the revised parameters on the planet properties. For example, \citet{Barclay2015} used NIR spectral indices and isochrone fitting to infer the individual stellar properties of KOI-1422 A+B, an M dwarf binary with five small transiting planets, while \citet{Gaidos2016} used photometry to classify and analyze a sample of \kep\ M stars, including some corrections for known binaries. \citet{Cartier2015} used high-angular resolution \textit{Hubble Space Telescope} imaging to measure the stellar properties of three \kep\ multiple systems by performing isochrone fitting. These valuable in-depth analyses of binary planet hosts have been time- and resource-intensive, and have been tailored to unique data sets and the needs of individual systems.

As the sample of binary planet host stars grows, such labor-intensive methods of characterization have become progressively less feasible, while the importance of considering planets in binary systems has only increased. Obtaining the data sets for in-depth analyses of binary systems remains labor-intensive, but for large samples with relatively homogeneous datasets like the \kep\ sample, it is useful to have a homogeneous analysis approach. One promising route forward is analysis of low-resolution spectra, which takes advantage of the increased information content provided by spectroscopy for relatively faint sources while simultaneously increasing accuracy over photometric methods. Using low-resolution spectroscopy, planets in binary systems can be characterized with high levels of precision at low observational cost, increasing the total sample of planets with accurate and reliable measurements. 

To leverage the strength of low-resolution observations of close binary star planet hosts, we have developed a Markov Chain Monte Carlo (MCMC) framework for measuring the stellar properties (temperature, surface gravity, radius, and distance) of the components of unresolved binary star systems. Because this algorithm uses unresolved, low-resolution spectra, the technique is not restricted to bright systems, relatively wide ($\rho > 1-2 \arcsec$) binaries, or sites with sub-arcsecond seeing, meaning that it is very flexible. Because it is automated, it has the capacity to analyze a large number of systems with minimal effort. In this paper we present our methods and an analysis of 8 K and M main sequence binary planet hosts with archival low-resolution optical spectra to validate our method and revise the parameters of an initial sample of binary planet hosts. Future papers in this series will apply this technique to new data.
 
\section{Sample Selection and Observations}\label{sec:sample} 
To test the derivation of parameters for unresolved binary components, we wished to identify \kep\ planet hosts that were binaries and had pre-existing, moderate-resolution, unresolved spectra. A sample where all observations came from the same instrument and observing configuration was also ideal, to facilitate the analysis process during these initial tests. One such sample is that of \citet{Mann2013}, hereafter M13, who observed a sample of 123 \kep\ K and M stars at moderate resolution (R $\sim 1200$) with the SuperNova Integral Field Spectrograph (SNIFS) at the University of Hawaii 2.2 meter telescope on Mauna Kea. The observations and data reduction are described in M13, and the reduced spectra are publicly available online\footnote{\url{https://cdsarc.unistra.fr/viz-bin/cat/J/ApJ/779/188}}. 

The SNIFS instrument is an integral field spectrograph with a 6$\times$6'' field of view covered by a 15$\times$15 grid of fibers, corresponding to a spaxel size of 0.4 arcsec$^{2}$ \citep{Aldering2002, Lantz2004}. SNIFS uses a dichroic to separate incoming light into red (5100$< \lambda < 9700$\AA, R $\sim$ 1000) and blue (3200 $< \lambda < 5300$\AA, R $\sim$ 800) arms \citep{Aldering2002, Aldering2006, Mann2012}. The M13 spectra were extracted and reduced using the standard SNIFS pipeline, which uses PSF fitting in each frame to extract the spectrum from the spectrophotometric data cube. 

We cross-referenced the M13 sample with that of \citet{Furlan2017}, who compiled high-resolution imaging results for \kep\ stars to identify close companions. We required spatially-resolved photometry to constrain the flux ratios (contrasts) between the binary components during the fitting process. In our final sample of targets, the compilations of contrasts in \citet{Furlan2017} were drawn from several different original sources. \citet{Law2014ApJ}, \citet{Baranec2016}, and \citet{Ziegler2017} presented results from the Robo-AO survey of \kep\ planet host stars, which used a custom filter \citep[LP600;][]{Baranec2014, Law2014SPIE} to perform speckle imaging. \citet{Horch2012} performed a Gemini North/DSSI speckle survey of \kep, CoRoT, and Hipparcos stars using the 562, 692, and 880 nm filters. \citet{Kraus2016} used Keck/NIRC2 with adaptive optics (AO) imaging and nonredundant aperture masking interferometry to identify binary companions to \kep\ planet hosts in the K band. Finally, \citet{Furlan2017} also included original observations from DSSI at Gemini North, the WIYN 3.5m, and the Lowell Discovery Telescope (speckle; 562, 692, and 880 nm) and used a combination of Keck/NIRC2 and instruments at the Palomar 5m, the Lick 3m, and the MMT to observe in the J and K bands. 

\begin{deluxetable*}{CCCCCCCC} 
\tablecaption{Target list, separations, and existing direct imaging contrasts}
\tablecolumns{8}
\tablewidth{0pt}
\tablehead{
\colhead{KOI} &
\colhead{Sep. (\arcsec)} &
\colhead{$\Delta m_{LP600}$} & \colhead{$\Delta m_{562 nm}$} & \colhead{$\Delta m_{692 nm}$}
& \colhead{$\Delta m_{880 nm}$} & \colhead{$\Delta m_{J}$} & \colhead{$\Delta m_{K_{p}}$}
}
\startdata
227 & 0.3 & 0.84 \pm 0.09 (1) & 1.03 \pm 0.15 (2) & 1.50 \pm 0.15 (2) & \nodata & \nodata & 0.02 \pm 0.01 (3)\\
1422 & 0.22 & \nodata & \nodata & 1.72 \pm 0.15 (5) & 1.62 \pm 0.15 (5) & 1.08 \pm 0.04 (2) & 1.16 \pm 0.02 (3)\\
1681 & 0.15 & \nodata & \nodata & 0.97 \pm 0.15 (2) & 0.40 \pm 0.15 (2) & \nodata & 0.069 \pm 0.016 (3)\\
2124 & 0.06 & \nodata & 0.51 \pm 0.15 (2) & \nodata & 0.18 \pm 0.15 (2) & \nodata & 0.01 \pm 0.01 (3)\\
2174 & 0.88 & 0.21 \pm 0.06 (1) & \nodata & \nodata & \nodata & 0.14 \pm 0.05 (2) & 0.10 \pm 0.01 (2) \\
2542 & 0.8 & 1.20 \pm 0.19 (1) & \nodata & \nodata & \nodata & \nodata & 0.932 \pm 0.002 (3)\\
2862 & 0.63 & 0.17 \pm 0.05 (4) & \nodata & \nodata & \nodata & \nodata & 0.00 \pm 0.01 (3)\\
3010 & 0.33 & \nodata & \nodata & 0.74 \pm 0.15 (2) & 0.01 \pm 0.15 (2) & \nodata & 0.25 \pm 0.05 (2)\\
\enddata
\tablecomments{(1) \citet{Baranec2016}; (2) \citet{Furlan2017}; (3) \citet{Kraus2016}; (4) \citet{Ziegler2017}; (5) \citet{Horch2012}}
\label{tab:targets}
\end{deluxetable*}

We selected known binary star planet hosts that were likely to have been unresolved in the M13 observations. The SNIFS instrument has a fiber size of 0.4", which means that all binaries closer than 0.2" should fall on the same spaxel. Whether or not a binary wider than 0.4" is resolved will be seeing-dependent: if the separation is less than the seeing on a given night, the combined light from the target will be modeled as a single source during the PSF fitting performed by the SNIFS pipeline during spectral extraction. We assumed a conservative seeing of 1", and so removed all targets from the cross-matched sample with a separation greater than 1". This left us with a list of 19 potential targets.

Finally, we investigated each target using the ExoFOP database \footnote{\url{https://exofop.ipac.caltech.edu/}} and removed any systems where there was more than one companion within 3", where the contrast was greater than 2.5 magnitudes (which corresponds to a minimum flux ratio of 0.1), or where the only planet candidate was flagged as a false positive. This was to streamline the fitting process, which becomes progressively more degenerate as the number of fit parameters increases. The limits of $\Delta$mag $< 2.5$ and $\rho<1\arcsec$ were imposed to ensure that some secondary star flux would be present in the spectrum, and to select for systems with real planet candidates. We also only selected systems that had more than one contrast available. At the end of this selection process, we were left with 8 targets for analysis, including one validation system, KOI-1422, that was studied in detail by \citet{Barclay2015}. The targets, their separation, and measured contrasts at various bandpasses from the direct imaging literature are all summarized in Table \ref{tab:targets}.

We visually inspected the high-resolution (speckle and AO) images uploaded to the ExoFOP for all systems to check for any inconsistencies between different identification and imaging methods. All of our systems were consistent with the multiplicity identified in \citet{Furlan2017}. We removed the $\Delta m_{[880]}$ contrast point for KOI-3010 from the analysis because it was discrepant with other contrasts, including a contrast in the \textit{Hubble Space Telescope} 775W band from \citet{Gilliland2015} ($\Delta m_{775W} = 0.294 \pm 0.05$ mag, while $\Delta m_{[880]} = 0.010 \pm 0.15$ mag). We removed the KOI-1681 $\Delta m_{[692]}$ contrast from the analysis because the secondary was not visible in the reconstructed image even though the reported contrast was $\Delta m_{[692]} = 0.97 \pm 0.15$ mag and the reported contrast lower limit was 6 mag at 0.2$\arcsec$, approximately the same separation as the binary companion.

\section{Spectral Fitting}\label{sec:analysis}
We had two categories of system: those with measured Gaia parallaxes and those without. We compiled a data set consisting of a single composite spectrum; a set of unresolved $JHK_{S}$ photometry from 2MASS \citep{Skrutskie2006} and $r'i'z'$ photometry from the KIC \citep{Brown2011}; and a set of contrasts from a variety of sources (Table \ref{tab:targets}). We did not include the $g'$ KIC photometry because spectral models often perform poorly below 5000 \AA. To retrieve the individual stellar parameters, we simultaneously fitted the total data set using synthetic contrasts, synthetic unresolved photometry, and a single model composite binary spectrum, all of which we generated using BT-Settl model spectra \citep{Allard2014, Baraffe2015} with the \citet{Caffau2011} line list\footnote{\url{https://phoenix.ens-lyon.fr/Grids/BT-Settl/CIFIST2011/SPECTRA/}}. To fit the 6 systems without a parallax, we also used all three data sets but neglected the final best-fit distance and absolute stellar radii and did not impose a distance prior. This was because the measured individual stellar temperatures were insensitive to distance errors, but the inferred absolute radii were sensitive to an unknown distance. However, the radius ratio was well-constrained using contrasts, so we were able to measure a radius ratio for all systems regardless of if they had a parallax. For both types of fit we implemented a modified Gibbs sampler algorithm to find the \cs\ minimum, then used \emcee\ to explore around the minimum and establish uncertainties.\footnote{Our fitting pipeline is available on GitHub: \url{https://github.com/kendallsullivan/binary_fitting_2022b}}

\subsection{System Model}
We assumed that our total data set (contrasts, unresolved photometry, and a spectrum) could be parameterized by 8 values: primary and secondary star effective temperatures, surface gravities, and radii, and system-wide extinction and distance/parallax. The temperatures and surface gravities impact the spectral lines; the radii change the relative normalization of the component spectra because the flux from each component is modulated by a factor of $R^{2}$; the radii and temperatures affect the contrasts; the distance, temperatures, extinction, and radii impact the unresolved photometry; and the temperatures and extinction alter the continuum slope of the spectrum. There is very little interstellar extinction in the Kepler field within a kiloparsec of the Sun \citep[e.g.,][]{Green2019}, so we fixed $A_{V} = 0$ for all fits.

There are other, less dominant factors like metallicity that we chose to neglect in our fitting for this study, but which could be added in the future. We neglected metallicity partially because the low spectral resolution of our data meant that we would not be able to measure metallicity with a high level of accuracy, and partly because stars in the Solar Neighborhood, such as our sample, typically have metallicity that is close to Solar \citep{Nieva2012, Bensby2014}. We tested a lower-metallicity ([Fe/H] = -0.5) model grid to explore the impact of lower metallicity on our results, but found that the differences between our results were negligible. We also could have allowed each star in a system to have its own distance, but because all the binaries in our sample are close binaries ($\rho \leq 1\arcsec$) that are highly likely to be bound \citep[e.g.,][]{Hirsch2017}, the differential distance between the two stars should be negligible.

Our goal was for our fitting method to be as model-independent as possible (excluding the reliance on stellar atmosphere models), so we chose not to anchor any parameters to an evolutionary model, meaning that the two stars were not required to have radii and temperatures that fell along the same isochrone. This increased the radius uncertainty but meant that our results were not tied to any particular set of physical assumptions, such as coevality. This is particularly important, for example, if a close binary is actually an unresolved hierarchical triple. In this scenario, the close unresolved pair could appear to have a larger radius than would be measured if the two observed sources were forced to fall on the same isochrone. By removing the model dependence, we eliminated these systematic errors, although we cannot identify outlier systems using only our analysis.

For systems with an observed parallax, we imposed a Gaussian prior on distance in parallax space using the measured parallax and associated error from \textit{Gaia} \citep{Gaia2016, Gaia2018, Luri2018}, then fitted for the primary radius, radius ratio of the system, and distance (in addition to temperatures and surface gravities). For systems without a parallax we only fit for the radius ratio of the system, and fit the unresolved photometry using an arbitrary distance which we optimized simultaneously with the other fit parameters.

\subsection{Spectrum Generation}
To find the best-fit set of parameters, we created synthetic data to compare with the observations. We had three different types of data to fit: a moderate-resolution optical spectrum, a set of measured contrasts, and a set of unresolved photometry. This subsection discusses the synthetic spectrum, while the following subsection outlines the synthetic photometry and contrasts, which were calculated using the synthetic spectra described here.

Briefly, we calculated an individual stellar spectrum using the proposed temperature and surface gravity for each component. We converted the synthetic stellar spectrum to the correct relative flux by either using the radius of each component and the system distance to infer a physical luminosity, or by normalizing the synthetic secondary spectrum luminosity relative to the primary luminosity using the radius ratio. Therefore, because the model spectra are in units of energy per unit area per unit wavelength, we multiplied each spectrum by (radius/distance)$^{2}$ to convert to a flux. For systems without a parallax, we approximated the parallax as the inverse of the KIC distance, and did not fit for the absolute radii, only the radius ratio of each system. Finally, we created the composite spectrum and normalized it so that the peak flux had a value of unity.

Our method for creating an individual stellar spectrum was described in detail in \citet{Sullivan2021}, and is briefly summarized here. We created a model spectrum for each stellar component by bilinearly interpolating BT-Settl model spectra \citep{Allard2014, Baraffe2015} with the CFIST 2011 line list \citep{Caffau2011, Allard2013, Rajpurohit2013} in temperature and surface gravity. We converted the high-resolution model spectra to the data resolution by convolving with a Gaussian corresponding to the instrumental spectral resolution, then downsampled the spectrum by interpolating the model spectrum onto the pixel scale of the data.

We did not impose stellar rotational velocity (v$\sin i$) or spatial radial velocity (RV) on the model spectra using the following logic: at the data spectral resolution of R$\sim$1000 and a wavelength of $\lambda_{cen} = 7500$ \AA, a typical stellar rotational velocity or spatial RV of 30 km/s would induce a wavelength broadening of $\Delta \lambda = 0.75$ \AA, which is an much smaller than either the $\sim 2$ \AA\ pixel size or the 7.5 \AA\ resolution element size. We also visually inspected the alignment between model and data spectral lines for each system, and did not find any apparent shifts larger than 50 km/s (1.25 \AA, or approximately half a pixel). Therefore we ignored the stellar rotational velocity or mutual spatial radial velocity when producing template spectra. After producing the individual stellar model spectra we calculated the contrasts, as described in the following subsection. After calculating the contrasts, we created a composite model spectrum by adding the two component spectra. 

\subsection{Photometry Generation}\label{subsec:phot}
We performed fits to two photometric data sets. First, we fit contrasts adopted from the high-resolution imaging surveys compiled by \citet{Furlan2017}. Second, we fit unresolved photometry in the $JHK_{s}$ 2MASS fiters \citep{Skrutskie2006} and SDSS $riz$ bands from the KIC \citep{Brown2011}.

After creating each individual spectrum and converting it to flux units from luminosity per surface area, we calculated contrasts in the relevant bands. Because contrast is a differential measurement, we did not need to use a filter zero point or otherwise calibrate the photometry to retrieve an flux that could be compared to observations. We convolved each model spectrum with the transmission curve for the relevant filter (typically obtained through the Spanish Virtual Observatory Filter Profile Service\footnote{\url{http://svo2.cab.inta-csic.es/svo/theory/fps3/index.php}}; \citealt{Rodrigo2012, Rodrigo2020}), then integrated over the curve to obtain an ``instrumental'' flux in each filter for each stellar component. The contrast was simply the difference between the primary and secondary magnitudes in each filter.

Finally, we calculated unresolved photometry in the KIC and 2MASS bands. We calculated the ``instrumental'' flux by convolving the composite model spectrum (in flux units) with the relevant filter, summing the flux across the bandpass, then dividing by the zero point of the filter before converting into a magnitude. For the KIC filters we calculated an AB magnitude in the available SDSS bandpasses, then converted to KIC AB magnitudes using the relations in \citet{Pinsonneault2012}. To match the 2MASS filters we calculated Vega magnitudes using the zero points from \citet{Cohen2003}.

\subsection{Chi-square Minimization to Identify Best-fit Parameters}
We first optimized our fit using an ensemble of modified Gibbs samplers, which is an algorithm that is frequently used for MCMC applications. The Gibbs sampler proposes a new set of test parameters and evaluates their quality, and usually accepts the new values if they are better, but sometimes does not. We modified our Gibbs sampler to always and exclusively accept proposals that resulted in a better fit (lower \cs\ value), to expedite the fitting. We chose to use modified Gibbs sampler rather than a simpler method like an amoeba fit to reduce the algorithm's sensitivity to local \cs\ minima, and instead of a more complicated method like gradient descent to avoid the complications introduced by differentiating an interpolated spectrum.

We initialized the fitting process with a set number of walkers, distributed in the 5 or 7-dimensional parameter space of (T1, T2, log(g)1, log(g)2, R2/R1, [R1, $\pi$]). All of the initial distributions were uniform across the parameter space, and we chose limits for each of the distributions based on the physical properties of our sample: for our sample of nearby M dwarfs, we typically imposed limits of $3000 < T < 5000$K, 4.5 $<$ log(g) $<$ 5.5, 0.05 \rsun$<$ R $<$ 1 \rsun, $0.05 < R2/R1$, and $1 < \pi < 10$mas, and we did not restrict the temperature ratio between the two components (i.e., the secondary was allowed to be fit as hotter than the secondary). We used 150 walkers at this initial stage to produce good initial parameter space coverage ($\langle \Delta T \rangle \sim 10$ K, $\langle \Delta \log(g) \rangle \sim 0.05$ dex, $\langle \Delta R \rangle \sim 0.005$\rsun). 

At each test step, all parameters were varied simultaneously by perturbing the old best fit value by a value drawn from a normal distribution. The mean of the distribution was the old best fit value, and the standard deviation varied depending on what stage of fitting the code was in. Initially, we had a large standard deviation, to search as much of parameter space as possible for each walker ($\sigma_{T} = 250$K, $\sigma_{\log(g)} = 0.1$,  $\sigma_{R2/R1} = 5\%$, $\sigma_{R}$ = 0.05 \rsun, $\sigma_{\pi} = 10\%$). Halfway through the fit, we switched to smaller values, assuming we had settled into a minimum on the \cs\ surface, to optimize the final value ($\sigma_{T} = 20$K, $\sigma_{\log(g)} = 0.05$, $\sigma_{R2/R1} = 1\%$, $\sigma_{R}$ = 0.01 \rsun, $\sigma_{\pi} = 5\%$). Our logic for choosing these values was similar to the reasoning described in \citet{Sullivan2021}. After perturbing the values, we generated a new data set (unresolved spectrum, contrasts, and unresolved photometry) using the methods described in the preceding subsections.

For each data set, we calculated the quality of the fit. To assess the quality of the photometry, contrast, and distance fits, we calculated the sum of the \cs\ for each data set. This provided a goodness-of-fit statistic encompassing $\sim$ 10 degrees of freedom (DoF). To assess the quality of the spectrum fit, we calculated the sum of the \cs\ values between the data and test spectra and added it to the total \cs\ after weighting by a factor $f_{spec}$. 

We imposed the weighting factor $f_{spec}$ using the following logic. The number of DoF of a spectrum is complex and difficult to assess, because there are covariances between different pixels of a spectrum \citep[e.g.,][]{Czekala2015}. However, to a first approximation, the DoF are the stellar temperature, surface gravity, extinction, and metallicity, with higher-order effects introduced by properties such as the specific chemical composition, the rotational velocity, and the magnetic activity of the star. The total number of spectrum pixels is $\sim 1200$ for our moderate-resolution spectra, which is much larger than the first-order 4 DoF of the spectrum. Therefore, if we were to simply calculate a reduced-\cs\ for the spectrum, the relative weight of the spectrum fit would be much larger than that of the other components of the goodness-of-fit assessment, meaning that the spectrum fit alone would drive the fit. 

To avoid these issues, we imposed the multiplicative reduction factor $f_{spec} = (n_{phot} + n_{contrast} + 1)/(n_{\text{res elements}})$ to down-weight the \cs\ contribution from the spectrum by dividing by the number of spectral resolution elements (i.e., the nominal reduced-\cs), then re-weight it using the total number of DoF from the rest of the data set. This down-weighting ensured that the contribution of \cs$_{spectrum}$ remained significant and more accurately reflected the $\sim 4$ DoF (to first order) of the spectrum. For example, for the test case of KOI-1422, $f_{spec} = (6 + 4)/400 \approx 0.025$, meaning that the sum of the calculated spectrum \cs\ values was multiplied by a factor of $\sim$0.025 before being added to the contrast \cs. This is analogous to calculating the reduced-\cs\ of the spectrum, then multiplying it by the total number of contrasts and photometric points, and reduces the spectral \cs\ from $\sim 10^{4}$ to $\sim 30$, which is comparable to the \cs\ contribution from the contrast calculation.

If a Gaia parallax for the system existed, we imposed a Gaussian prior on the parallax, which we added to the \cs\ with the same weight as the spectroscopic, photometric, and contrast components. Thus, for systems with a Gaia parallax, the total \cs\ equation was $\chi^{2} = f_{spec}\chi^{2}_{spec} + \chi^{2}_{phot} + \chi^{2}_{contrast} + \chi^{2}_{distance}$. For systems without a Gaia parallax, the \cs\ equation was simply $\chi^{2} = f_{spec}\chi^{2}_{spec} + \chi^{2}_{phot} + \chi^{2}_{contrast}$.The reduced \cs\ of the spectrum term was typically larger than 1, while the reduced \cs\ of the contrast and photometry terms was $\sim$1. However, the spectral reduced \cs\ is larger than 1 because of model spectrum properties that affect every proposed model similarly (e.g., poorly modeled spectral bands), rather than being large as a result of a badly-fitting underlying system model (T1, T2, R1, R2/R1, etc). Although the magnitude of the \cs\ is greater than the number of degrees of freedom, the differential \cs\ (i.e., the difference between a better or worse fit) should not be affected by these modeling systematics, but instead should remain sensitive to the intrinsic properties of the spectrum (the underlying system model). There is a more statistically robust way to combine multiple likelihoods \citep[e.g.,][]{Czekala2015}, but it is computationally infeasible and this method produces satisfactory results in more practical run times.

If the calculated \cs\ was smaller than the previous best guess, the guess was accepted and the step counter was reset to zero, while if the fit was worse the guess was discarded and the step counter was increased by one. We ran the optimization stage until each walker had individually made 600 guesses without finding a better match. 

\subsection{MCMC Determination of Parameter Uncertainties}
At the end of the optimization stage, we recorded the best \cs\ value and corresponding set of fit parameters for all walkers. We selected the 30\% of walkers with the lowest \cs\ values and used them as the initial positions to initialize \emcee, a Bayesian MCMC sampler \citep{DFM2013}, to assess the errors on each of the fit parameters. We again imposed a Gaussian prior for parallax from \textit{Gaia} when available, and uniform priors otherwise. We evaluated a likelihood function that assessed the reduced-chi square value for the total data set in the same manner as described above for the optimization step. 

We ran \emcee\ until it converged or reached 15,000 steps, whichever came first. We assessed convergence using convergence criteria based in the autocorrelation time following the prescription of \citet{DFM2013}. Using these criteria, we asserted that convergence was reached when the length of the chain is $\sim$50 times longer than the autocorrelation time $\tau$, and the fractional change in $\tau$ between steps is less than 10\% (i.e., ($\tau_{N} - \tau_{N-1}$)/$\tau_{N} < 0.1$). If the posteriors were bimodal with widely separated modes, we fit a double-peaked Gaussian distribution and took the parameters of the curve with the largest area to be the best-fit values. In both cases the chosen mode was also more consistent with the expected values for a main-sequence star, and the \cs\ value for the chosen mode was smaller than the \cs\ for the smaller peak.

\section{Test System Results}\label{sec:test results}
In this section we summarize several spectral analysis test cases: GJ 544B, a single star to verify our technique; a suite of synthetic unresolved binary spectra; and KOI-1422, which was studied in detail by \citet{Barclay2015} and was included in the M13 sample. 

\subsection{Single Star Results} 
We began by testing our fitting technique on a single star (GJ 544 B) included in \citet{Mann2015}, hereafter M15. The temperature measurements from that study were calibrated using stars that had interferometric radius measurements, so their technique was vetted using an additional constraint beyond simply comparing to model spectra. M15 generally found a temperature measurement disagreement of $<$ 20 K between their model-derived temperatures and the interferometric temperatures. Thus, we view their results as being more reliable than if they had only used model fitting to derive temperatures, and use one of their single stars as an initial test of our retrieval capabilities. M15 measured a temperature of \teff = 3191 $\pm$ 60 K for GJ 544B, where the 60 K error was assessed by comparing the variance between temperature assessments from different models and including a calibration error term as measured in M13.

\begin{figure}
\plotone{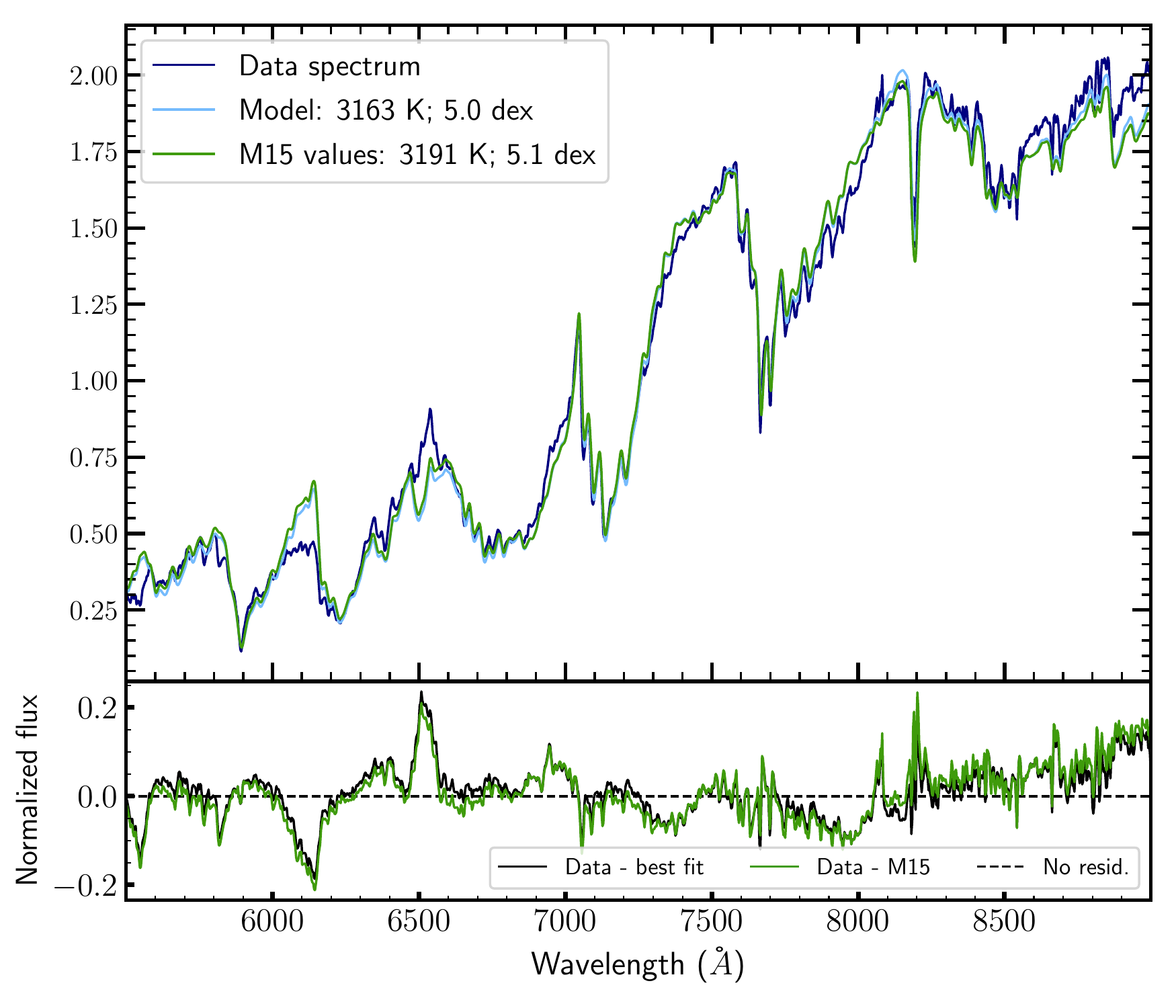}
\caption{The best fit spectrum (cyan), a spectrum created with the best-fit parameters from M15 (green), and the SNIFS spectrum (blue) for GJ 544B after the initial optimization process. The best-fit parameters from this work differ slightly from the M15 results but the resulting spectra are very similar matches to the data. The bottom panel shows the residuals for the M13 best fit (green) and our best fit (black). The residuals for the two fits are nearly identical, and are typically small except in some poorly-modeled spectral features.}
\label{fig:single spec}
\end{figure}

\begin{figure}
\plotone{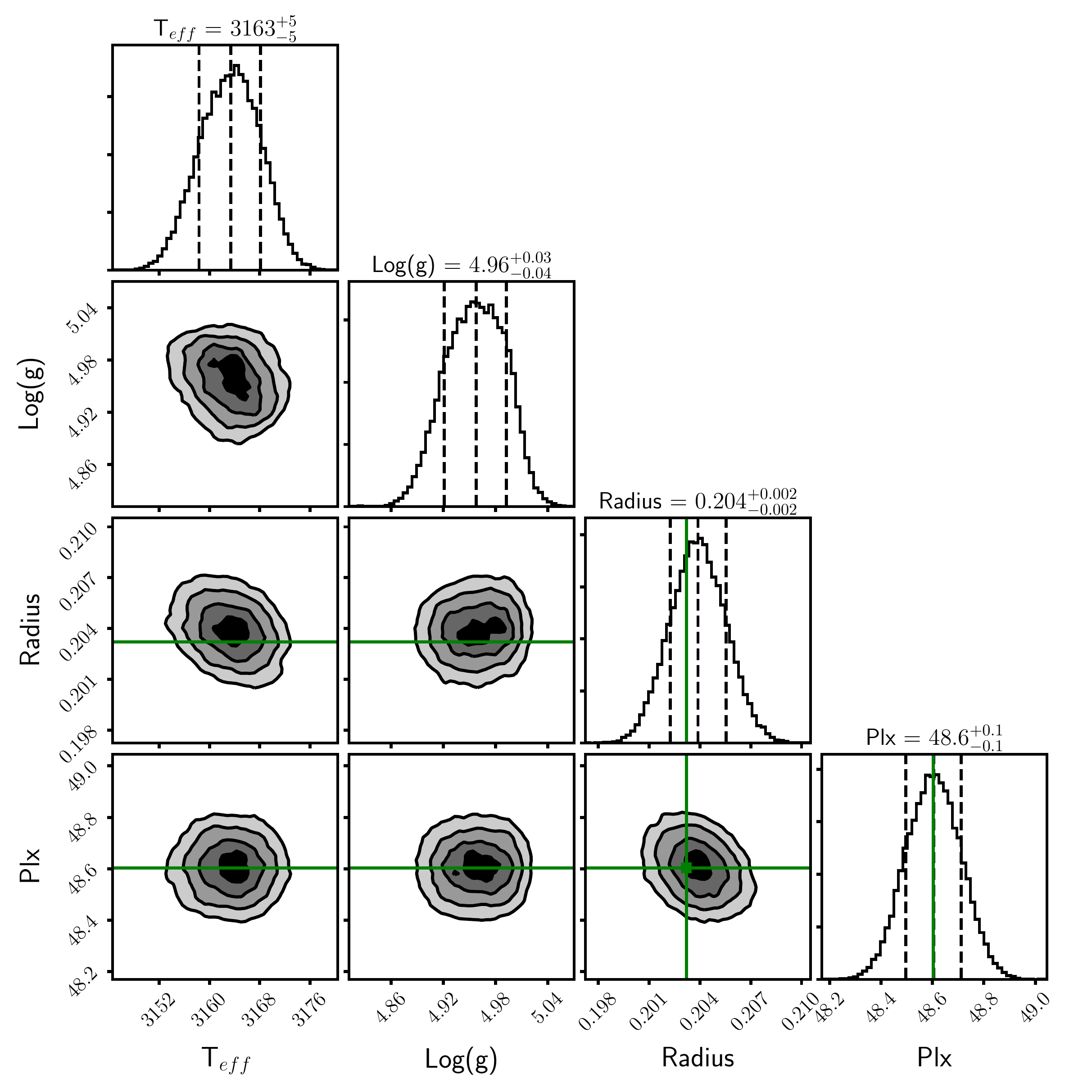}
\caption{The marginalized posterior distributions for the parameters of GJ 544 B. The green lines show the values measured by M15 and \textit{Gaia}. The radius and parallax are recovered within error, and the temperature is retrieved within the 60 K error of the M15 measurement. The surface gravity is not retrieved within error.}
\label{fig:single corner}
\end{figure}

We fit the spectrum using the BT-Settl models to match M15 and used our binary fitting technique after modifying it to accommodate a single star by reducing the number of fit parameters to be a single temperature, radius, surface gravity, and distance. Because the distance of GJ 544 B is only 20 pc ($\mu = 48.604 \pm 0.095$ mas; \citealt{Gaia2018}), we assumed extinction was negligible and so fixed \av = 0, which is the same assumption made by M15. We fit the SNIFS spectrum using only the red arm (5100-9000 \AA), because the SNIFS throughput drops dramatically in the blue. 

The best-fit model spectrum after the optimization stage is plotted alongside the input data and best-fit model from M15 in Figure \ref{fig:single spec}, and the covariances and marginalized posterior distributions from the MCMC error analysis are visualized in Figure \ref{fig:single corner}.

We measured a temperature of 3163$\pm$5 K, which is consistent with the M15 measurement of 3191$\pm$60 K. The error of the MCMC fit is much smaller than the M15 result because of the different methods of error calculation: the M15 error is driven by systematic error in the model atmosphere grids, whereas the error calculated by our MCMC is statistical, and so does not capture the grid-to-grid systematic uncertainty in temperature scales. The measured radius of $0.204\pm 0.002$ \rsun is also consistent with the previously measured radius of 0.20$\pm$0.01\rsun. The error on the radius is driven by the error on the parallax, and so is an order of magnitude smaller than the model-to-model uncertainty quoted by M15. The measured mass and radius of GJ 544 B from M15 correspond to an approximate surface gravity of $\log(g) = 5.11\pm 0.04$ dex, which is discrepant with our measurement of $\log(g) =$\val{4.96}{0.03}{0.04} dex. We conclude that the fitting algorithm recovers stellar temperatures and radii accurately enough to be extended to binary stars, but that the surface gravity measurements may not be reliable.

There are several spectral regions with significant mismatch between the data and the best-fit model. M13 observed similar mismatches to their spectra using the same BT-Settl model suite, and compensated by masking spectral regions with the largest systematic discrepancy while fitting. We chose to include those spectral regions rather than masking them, because of the relatively small number of total pixels in each spectrum, and because different spectral types would require different regions to be masked. The discrepancy is due to the difficulty of modeling stellar atmospheres, especially in cool stars, and thus is a limitation of using stellar models instead of observed reference spectra to fit the data. These large regions of mismatch span tens of pixels and are evidently covariant, and so demonstrate the necessity of our procedure down-weighting the spectral \cs\ while fitting to the observed spectrum.

\subsection{Synthetic Spectrum Results}\label{subsec:synth data}
\begin{deluxetable*}{CCCCCCCCCCC}
\tablecaption{Synthetic Data Parameters \label{tab:synth params}}
\tablecolumns{11}
\tablewidth{0pt}
\tablehead{
\colhead{\teff$_{,1}$} &
\colhead{R$_{1}$} & \colhead{log(g)$_{1}$} & \colhead{log(L)$_{1}$} & \colhead{\teff$_{,2}$} & \colhead{R$_{2}$} & \colhead{log(g)$_{2}$} & \colhead{(log(L)$_{2}$)} & \colhead{R$_{2}$/R$_{1}$} & \colhead{$\Delta m_{r'}$} & \colhead{$\Delta m_{K}$}\\
\colhead{(K)} & \colhead{(\rsun)} & \colhead{(dex)} & \colhead{(log(L$_{\odot}$))} & \colhead{(K)} & \colhead{(\rsun)} & \colhead{(dex)} & \colhead{(log(L$_{\odot}$))} & \colhead{} & \colhead{(mag)} & \colhead{(mag)}
}
\startdata
\multirow{5}{*}{3850} & \multirow{5}{*}{0.4994} & \multirow{5}{*}{4.76} & \multirow{5}{*}{-1.309} & 3025 & 0.1546 & 5.16 & -2.746 & 0.31 & 5.30 & 3.28\\
& & & & 3225 & 0.2149 & 5.06 & -2.349 & 0.43 & 3.69 & 2.39\\
& & & & 3425 & 0.3048 & 4.96 & -1.941 & 0.61 & 2.24 & 1.44\\
& & & & 3625 & 0.3910 & 4.87 & -1.626 & 0.78 & 1.11 & 0.71\\
& & & & 3800 & 0.4745 & 4.79 & -1.376 & 0.95 & 0.23 & 0.15\\
\hline
\multirow{6}{*}{4200} & \multirow{6}{*}{0.6160} & \multirow{6}{*}{4.67} & \multirow{6}{*}{-0.9546} & 3225 & 0.2149 & 5.06 & -2.349 & 0.35 & 4.85 & 3.08\\
& & & & 3425 & 0.3048 & 4.96 & -1.941 & 0.49 & 3.40 & 2.13\\
& & & & 3625 & 0.3910 & 4.87 & -1.626 & 0.63 & 2.26 & 1.41\\
& & & & 3825 & 0.4870 & 4.77 & -1.342 & 0.79 & 1.27 & 0.77\\
& & & & 4025 & 0.5791 & 4.70 & -1.103 & 0.94 & 0.47 & 0.24\\
& & & & 4175 & 0.6039  & 4.68 & -1.003 & 0.98 & 0.08 & 0.06\\
\enddata
\tablecomments{All system parameters were calculated using a MIST isochrone at an age of 5 Gyr.}
\end{deluxetable*}

To establish the potential limits of the fitting algorithm, we simulated a series of unresolved binaries with different temperature/flux ratios and primary temperatures, and fit them to retrieve the input parameters. We simulated systems with primary stars at temperatures of 3850 K and 4200 K, with secondary stars of various temperatures ranging from 3025 K to 3800 K for the cooler primary and 3225 K to 4175 K for the hotter primary. Other system properties such as the component surface gravities, radii, and luminosities, were calculated using the MESA Isochrones and Stellar Tracks (MIST) isochrones \citep{Paxton2011, Paxton2013, Paxton2015, Dotter2016, Choi2016} at an age of 5 Gyr. These parameters, as well as the r' and K band contrasts for each system, are listed in Table \ref{tab:synth params}. For each system we generated a composite spectrum, synthetic photometry in the $riz$ and JHK$_{s}$ bandpasses, and $\Delta m{[880]}$, $\Delta m_{J}$, and $\Delta m_{Kp}$ contrasts. We added Gaussian noise at the 1\% level to each spectrum to mimic the fitting process in the case of high-quality (SNR $\sim$ 100) data. We also assumed $\sigma_{phot} = 0.05$ mag and $\sigma_{\Delta} = 0.02$ mag, to match the typical precision of the real data.

We analyzed each system using the methodology described above in Section \ref{sec:analysis}, assigning each generated spectrum an arbitrary distance of 100 $\pm$ $\sim$ 10 pc ($\mu = 0.01\pm0.001\arcsec$) when calculating the input ``measured'' photometry. Our assumed error of $\sim$ 10\% was comparable to the largest parallax error in our sample. When calculating the fit photometry we fit in the manner described for systems without a measured parallax by assuming an arbitrary distance and only measuring the resulting radius ratio, rather than imposing a prior on the distance and measuring the absolute stellar radii. We did not impose a radial velocity (RV) shift between the two spectral components, because typical RV shifts of 50 km/s for close binaries induce wavelength shifts of only 1.25\AA, which is approximately half the pixel width and so would not be resolved in the spectrum.

\begin{figure*}
\gridline{\includegraphics[width = 0.45\linewidth]{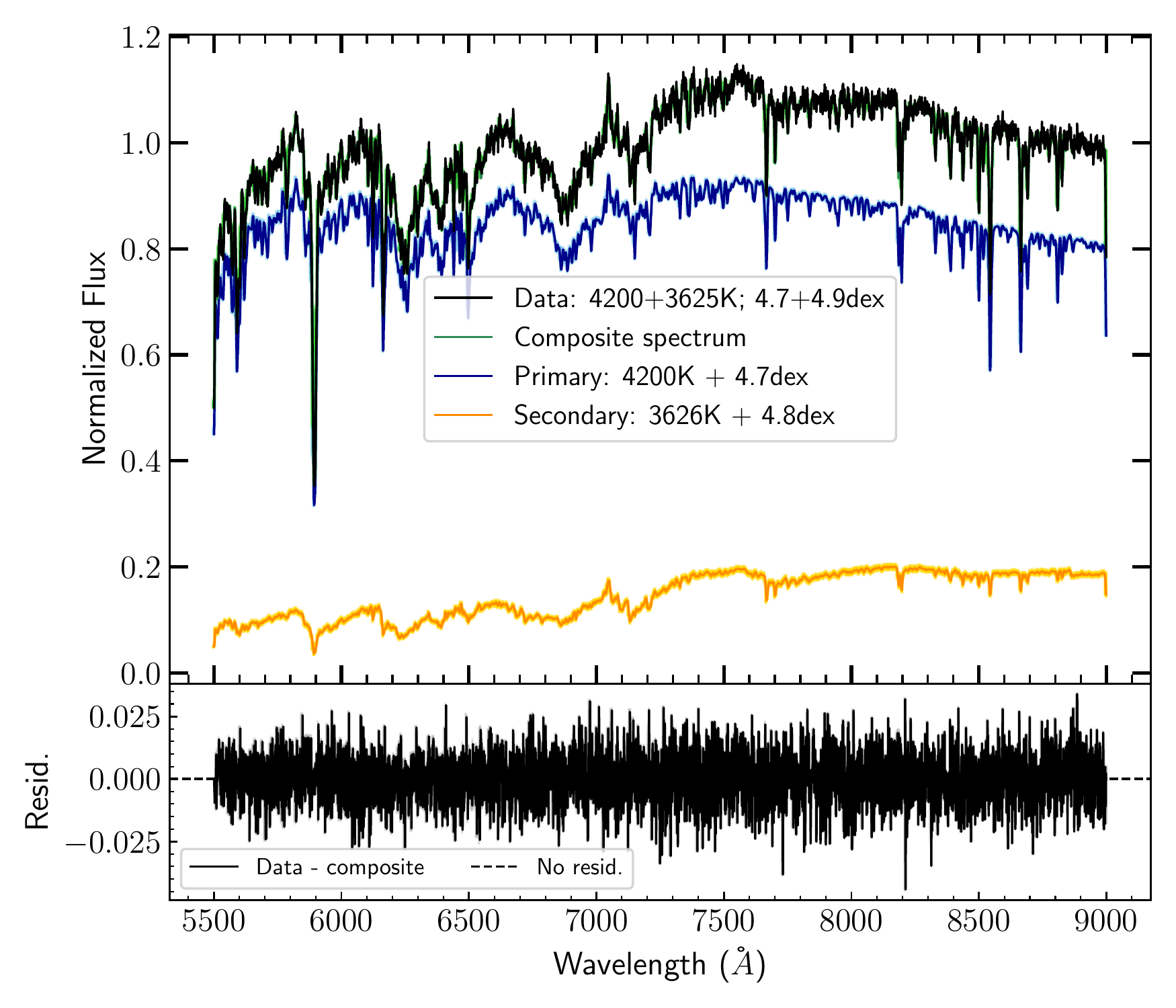}
	\includegraphics[width = 0.45\linewidth]{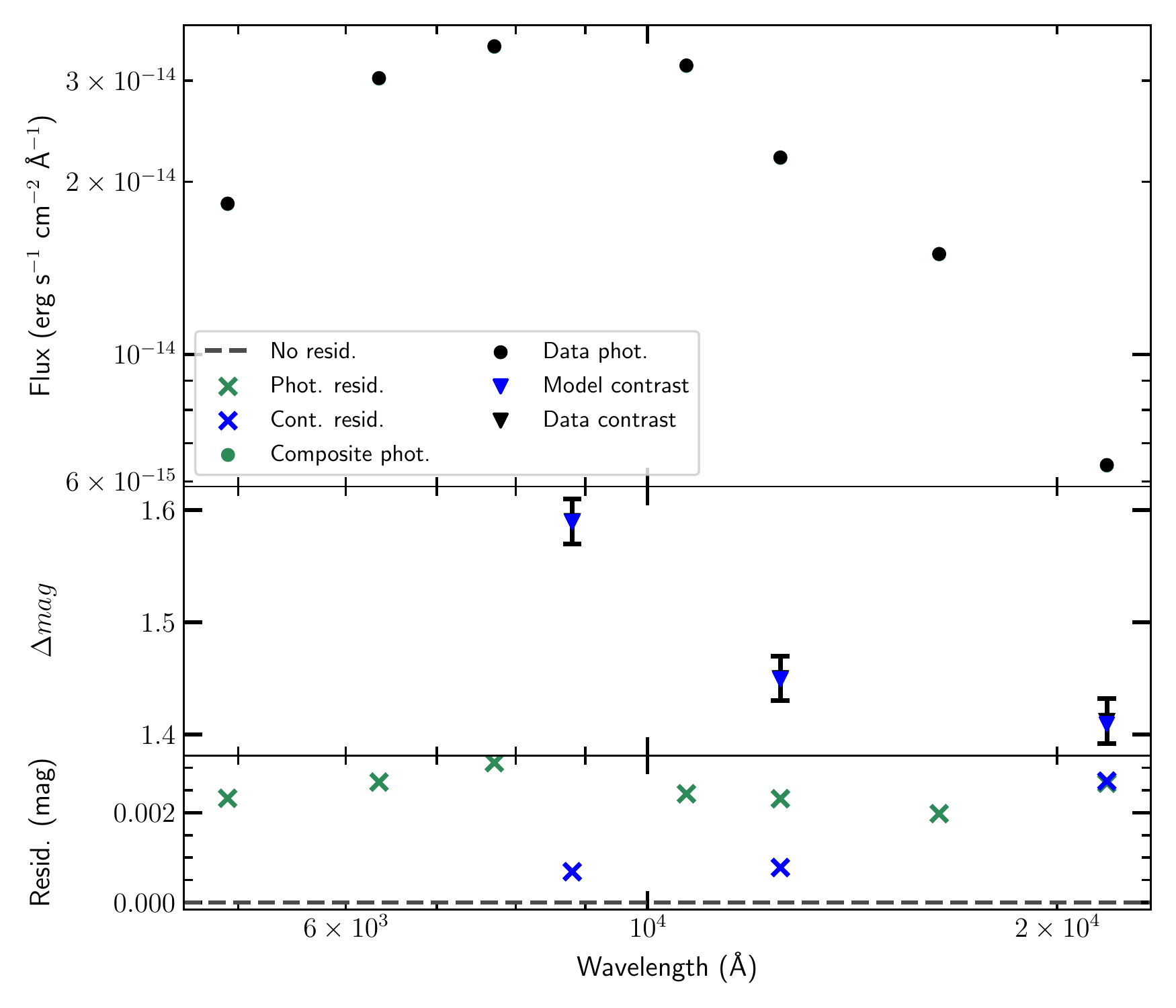}
}
\caption{Example fit for a synthetic binary. Left: The data spectrum (black), best-fit spectrum (green), and primary and secondary component spectra (blue and gold, respectively). 100 random samples from the MCMC chains are plotted for the three model spectra, but are not obviously apparent because of the tightly constrained fit. The bottom panel shows the residual between the data and the best-fit model (black) and the 100 random draws (gray underlay). The model is a good fit to the synthetic data, which is evident because the model is not visible under the data spectrum. The residuals are on the order of 1\%, consistent with the 1\% Gaussian noise imposed on the synthetic data. Right: The top panel shows the ``observed'' (black points) and best-fit model (green points) photometry. The middle panel shows the observed (black triangle) and best-fit model (blue triangle) contrasts. The bottom panel shows the residuals for both the photometry and the contrasts using green and blue `x' markers, respectively. The residuals are $\lesssim$ 0.003 mag in every case, indicating that the fit to the photometry is very good.}
\label{fig:synth spec}
\end{figure*}

Figure \ref{fig:synth spec} shows representative fit summary plots from one of the analyses of synthetic data (T1 = 4200 K, T2 = 3625 K). The left panel shows the data spectrum (black), the best-fit model spectrum (green), and the two best-fit component spectra (blue and gold for the primary and secondary stars respectively). 100 random samples from the MCMC chains are plotted underneath the three model spectra, but the best-fit values are so closely constrained that they are not apparent. Similarly, the best-fit model is a good enough fit to the data that it is not evident in the figure. The bottom inlay of the left panel shows the residuals for the best-fit spectrum and the 100 random draws in black and gray, respectively. The residual is typically on the order of 1-2\%, which is consistent with the artificial 1\% Gaussian noise imposed on the synthetic data. The right panel of Figure \ref{fig:synth spec} shows the observed and best-fit photometry (top panel), contrasts (middle panel), and residuals (bottom panel). The residuals are less than 0.02 mag for both photometric data sets, which corresponds to a discrepancy of $\sim$ 2\% at most.  

\begin{figure*}
\plotone{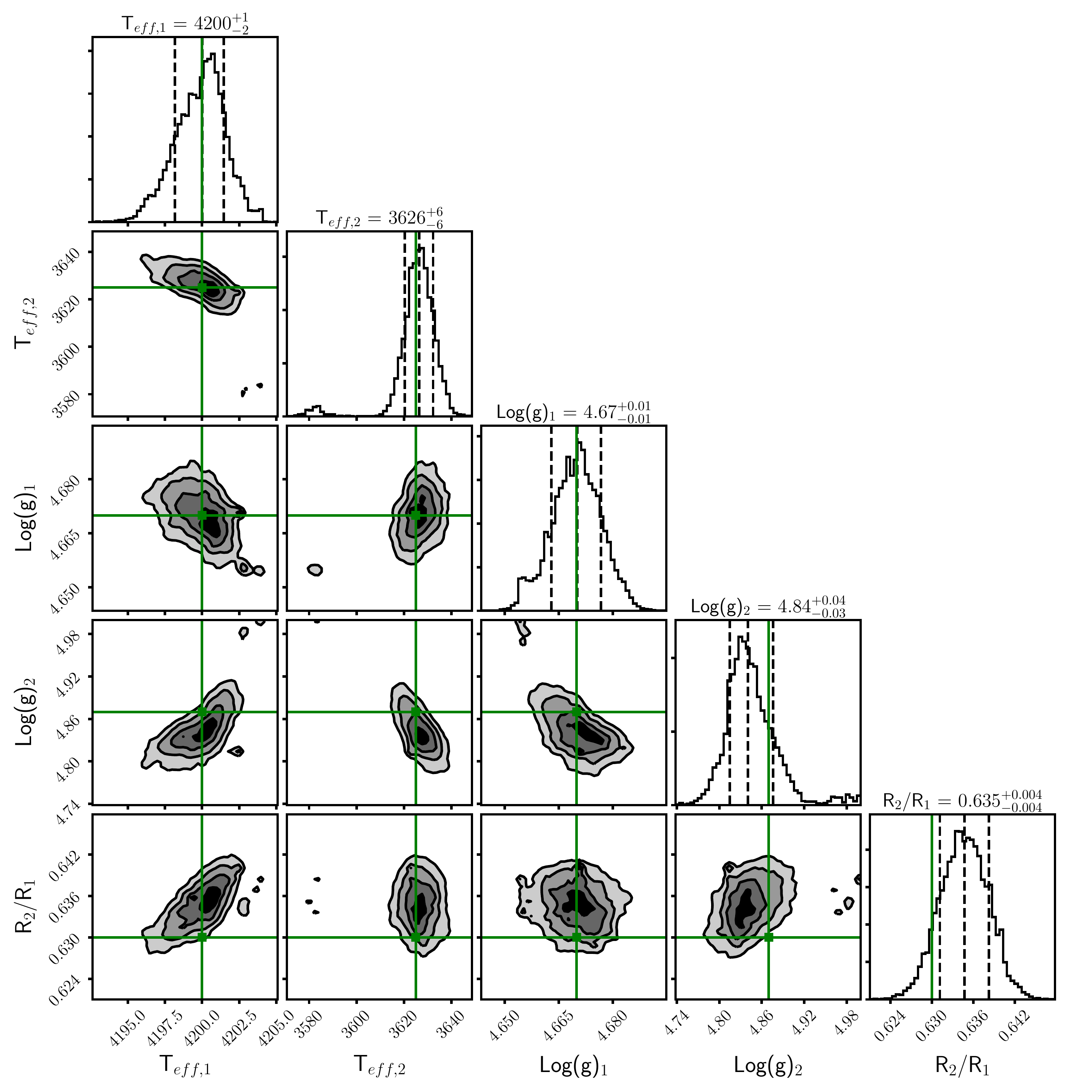}
\caption{An example of a corner plot showing the marginalized posterior distributions and covariance plots for a sample synthetic binary. The green lines on each plot show the true values for each parameter.}
\label{fig:synth corner}
\end{figure*}

The marginalized posterior distributions and covariance plots for the example synthetic data test system are shown in a corner plot in Figure \ref{fig:synth corner}. We measured temperatures of T1 = \val{4200}{1}{1}K, T2 = \val{3626}{6}{6}K, log(g)1 = \val{4.67}{0.01}{0.01}, log(g)2 = \val{4.84}{0.04}{0.03}, R2/R1 = \val{0.63}{0.005}{0.005}. The true values for the system are T1 = 4200 K, T2 = 3625 K, log(g)$_{1}$ = 4.67, log(g)$_{2}$ = 4.87, and R2/R1 = 0.63. Thus, all the parameters for this simulated system were recovered correctly within the statistical error.

\begin{figure*}
\gridline{\includegraphics[width = 0.45\linewidth]{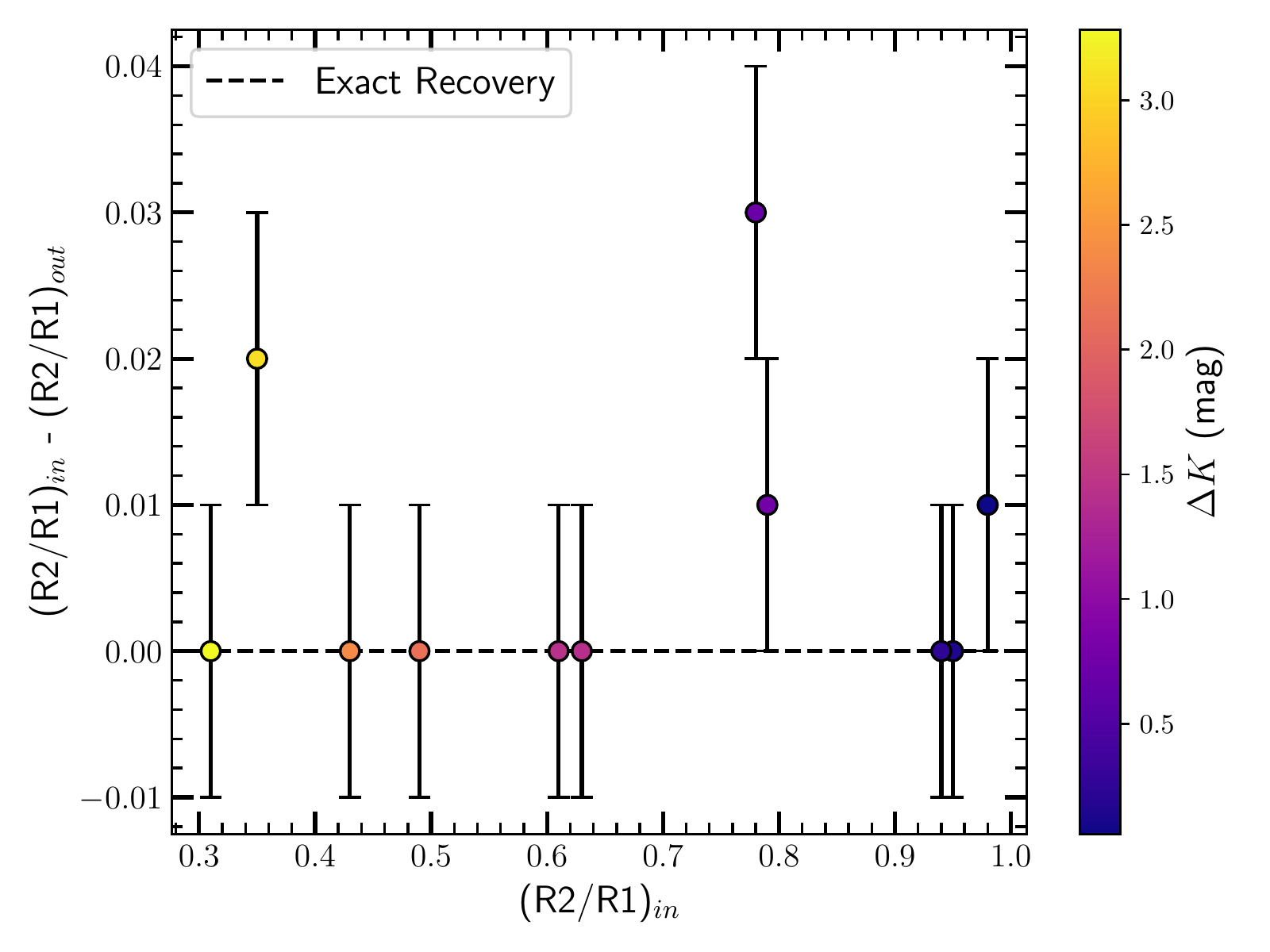}
	\includegraphics[width = 0.45\linewidth]{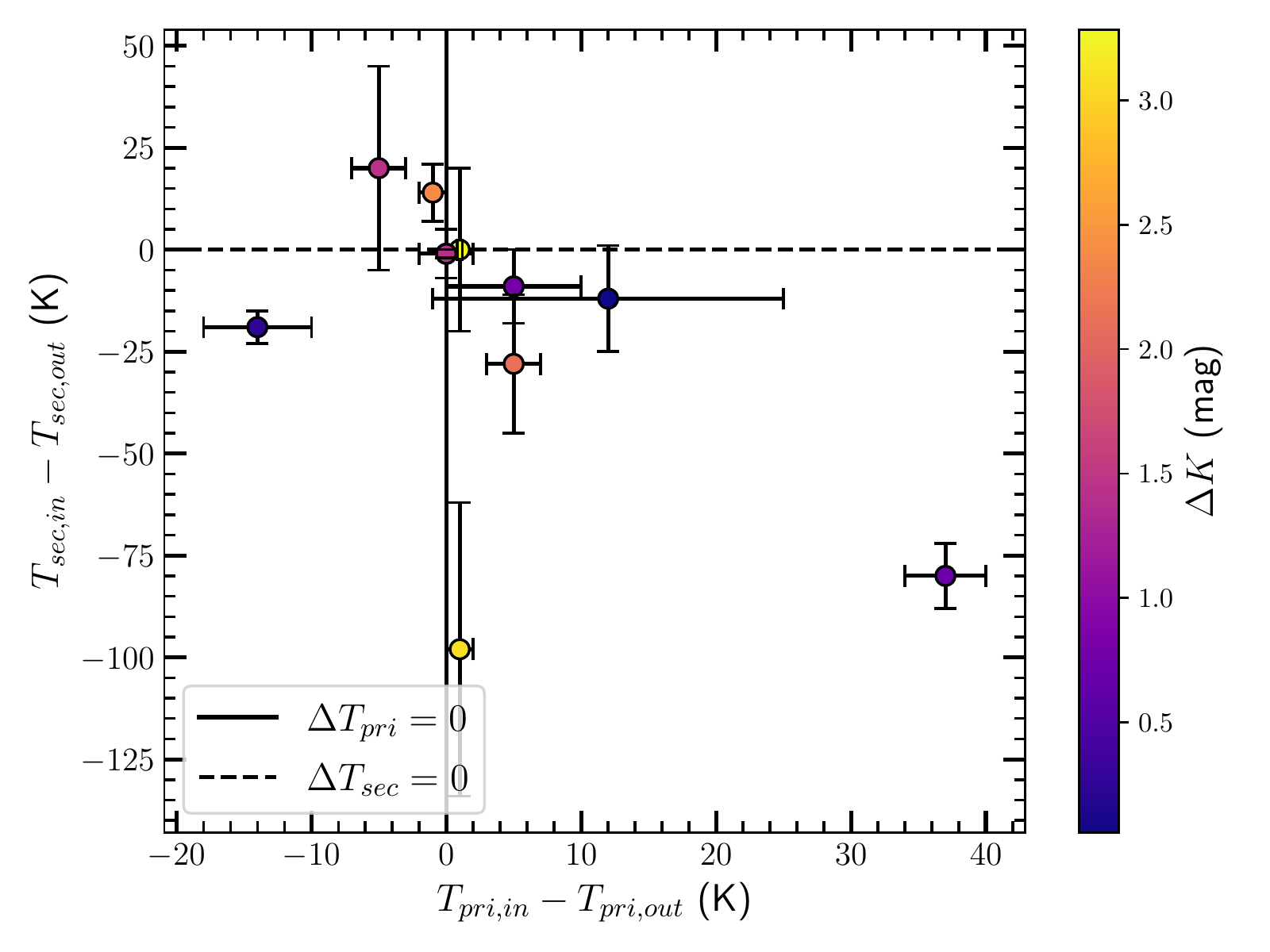}
}
\gridline{
\centering
\includegraphics[width = 0.45\linewidth]{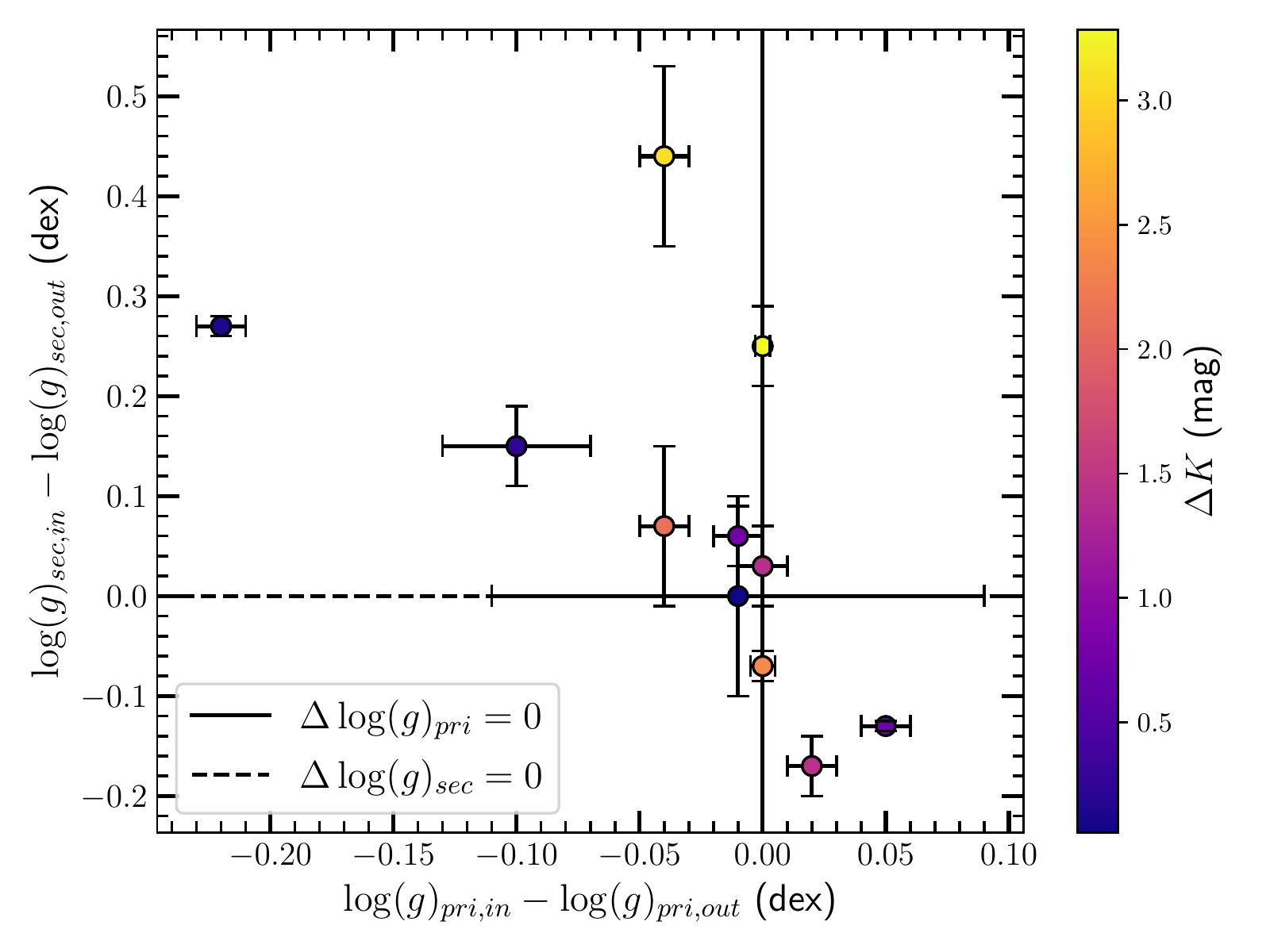}
}

\caption{Synthetic data fit results. Top Left: The radius ratio residual (input - output ratio) for each system plotted against the input (known) initial radius ratio, color-coded by the $K$-band contrasts between the primary and secondary stars. Most systems have the radius ratio recovered within 0.01 R2/R1. Top Right: The secondary temperature residual plotted against the primary temperature residual for the synthetic systems, color-coded by the $K$-band contrasts between the two stars. The majority of systems have the primary temperature recovered within 10 K and the secondary temperature recovered within 25 K. The decreased accuracy for the secondaries is because of their smaller contribution to the total flux in the spectrum. The systems that are significant outliers typically have large $K$-band contrasts between the components. Bottom: The secondary surface gravity residual plotted against the primary surface gravity residual, color-coded by the $K$-band contrasts between the two stars. The primary star surface gravity is typically recovered within 0.1 dex, but the secondary star surface gravity is often more than 0.1 dex discrepant. The quality of the retrieval does not appear to depend on the contrast between the components. }
\label{fig:synth summary}
\end{figure*}

Figure \ref{fig:synth summary} shows summary plots for the results from the synthetic spectra we analyzed. The top left panel shows the radius ratio residual plotted against the input radius ratio, with each point color coded by the $K$-band contrast between the two components. The simulated systems typically had radius ratios that were measured to within $\Delta_{R_{2}/R_{1}} = 0.02$. The two most significant radius ratio outliers, with residuals around $\Delta R_{2}/R_{1} \sim$ 0.03, were both systems with small-to-intermediate radius ratios and moderate-to-large $\Delta K$ between the two components, both of which are factors that significantly reduce the flux contribution from the secondary star within the data set. However, even in those most extreme cases, the radius ratio error was at most $\sim 5\%$. 

The top right panel of Figure \ref{fig:synth summary} shows the temperature residuals for the secondary stars plotted against the temperature residuals for their primary stars, with each point color coded by the $\Delta K$ between the primary and secondary star. For all systems, regardless of contrast, the primary star temperature is recovered to within $\sim$35 K, while the majority of the secondary star temperatures are recovered to within 30 K. The system with the largest secondary star residual of $\sim$ 100 K has a primary temperature that is recovered almost exactly and is a high-contrast system, indicating that the temperature discrepancy is a result of minimal flux being contributed to the composite spectrum by the secondary star. The system with the largest $\Delta K$ has a secondary temperature that is recovered more accurately because the primary star has very shallow molecular TiO bands while those of the secondary star are very deep, making it easier to identify the properties of the secondary even though it is contributing a very small amount of flux to the total spectrum. 

The other most significant outlier is a system with a moderate flux ratio, with T$_{1}$ = 3850 K, T$_{2}$ = 3625 K, but recovered values of T$_{1}$ = 3815 K and T$_{2}$ = 3705 K. There is a degeneracy between primary and secondary star temperature recovery, where a too-cool primary star temperature measurement can be complemented by a too-warm secondary star temperature. In the extreme flux-ratio cases, where R$_{2}$/R$_{1}\rightarrow$ 1 or R$_{2}$/R$_{1}\rightarrow$ 0, the fit is driven almost entirely by the contrasts. For example, in the largest flux-ratio case where $\Delta r = 5.30$ (T$_{1}$ = 4200 K, T$_{2}$ = 3225 K), the secondary star's contribution to the spectrum is on the order of the 1\% Gaussian noise imposed on the model spectrum, and so the recovery of the secondary star temperature is driven purely by the best-fit contrasts. Similarly, for the largest flux-ratio cases, where $\Delta r \sim 0.05$, the spectrum is not informative because the two spectra are approximately identical, again meaning that the temperature recovery is driven by the best-fit contrasts. In the intermediate regime, where the contrast is $\Delta r \sim 1$ mag (T$_{1}$ = 3850 K, T$_{2}$ = 3625 K), the temperature difference is adequate to allow the spectrum to begin to inform the fit alongside the contrasts, meaning that the degeneracy between primary and secondary star temperatures when measured using a spectrum comes into play and introduces larger offsets in the temperature recovery. 

However, even in the least accurate fit, the temperature offsets are $\Delta T_{1} \sim 40$ K, $\Delta T_{2} = 80-100$ K, which are within the typical errors of temperature measurements for single stars. Furthermore, at a temperature of 3500 K and an age of 5 Gyr, a temperature change of $\sim$ 40 K corresponds to a brightness change of $\Delta K \sim 0.02$ mag, which means that the most discrepant primary star temperature measurement is also consistent with the error we assumed for the photometry. Similarly, the secondary star is 1 mag, or a factor of $\sim$3, fainter than the primary, so the corresponding expected temperature error is $\sim\sqrt{3}$ larger than that of the primary, or $\sim$70 K, which is consistent with the secondary star discrepancy.

The bottom panel of Figure \ref{fig:synth summary} shows the secondary star surface gravity residual plotted against the primary star surface gravity residual. The primary star surface gravities are typically retrieved fairly well, with only one system more than 0.1 dex discrepant, and most systems consistent within 0.05 dex. The secondary star surface gravities often disagree with the true value by 0.1 dex or more. The expected range of surface gravity for main sequence M stars is $\sim 4.9 \pm 0.14$ (taken as the mean and standard deviation of our simulated sample), meaning that the observed large deviations from the secondary star log(g) indicate that the surface gravity for secondary stars is poorly constrained.

From this analysis of 11 synthetic binary systems, we conclude that for our sample of M stars with moderate contrasts and high-quality spectra we should be able to successfully recover the radius ratios and the component temperatures for each system with minimal measurement error. We were not able to accurately recover most surface gravities for secondary stars, so we anticipate that we will not be able to measure surface gravity in real data.

\subsection{Previously Analyzed Data: KOI-1422}
To validate our technique by comparing our results with those of a different method of deriving individual stellar and planetary properties, we analyzed KOI-1422 A+B. KOI-1422 is a \kep\ M dwarf binary with components separated by $0\arcsec.22$ that hosts 5 small transiting planets, and was previously studied by \citet{Barclay2015}, hereafter B15. The B15 analysis used NIR spectral indices, a method developed by \citet{Covey2010} and \citet{Rojas-Ayala2012}, and evolutionary model fitting to determine the properties of the two stars in the system using J and K band contrasts. KOI-1422 A+B was also observed by M13, meaning that there is an unresolved SNIFS spectrum of the system. B15 measured values of T$_{1}$ = 3740$\pm$130 K, T$_{2}$ = 3440$\pm$75 K, log(g)$_{1}$ = \val{4.77}{0.09}{0.06}, log(g)$_{2}$ = \val{4.93}{0.09}{0.06}, R$_{2}$/R$_{1}$ = \val{0.67}{0.09}{0.11}. The M13 temperature for the system was \teff = 3622$\pm$58 K, which is a 100 K deviation from the true primary temperature and $\sim$ 200-300 K hotter than the primary star temperature.

\begin{figure*}
\gridline{\includegraphics[width = 0.45\linewidth]{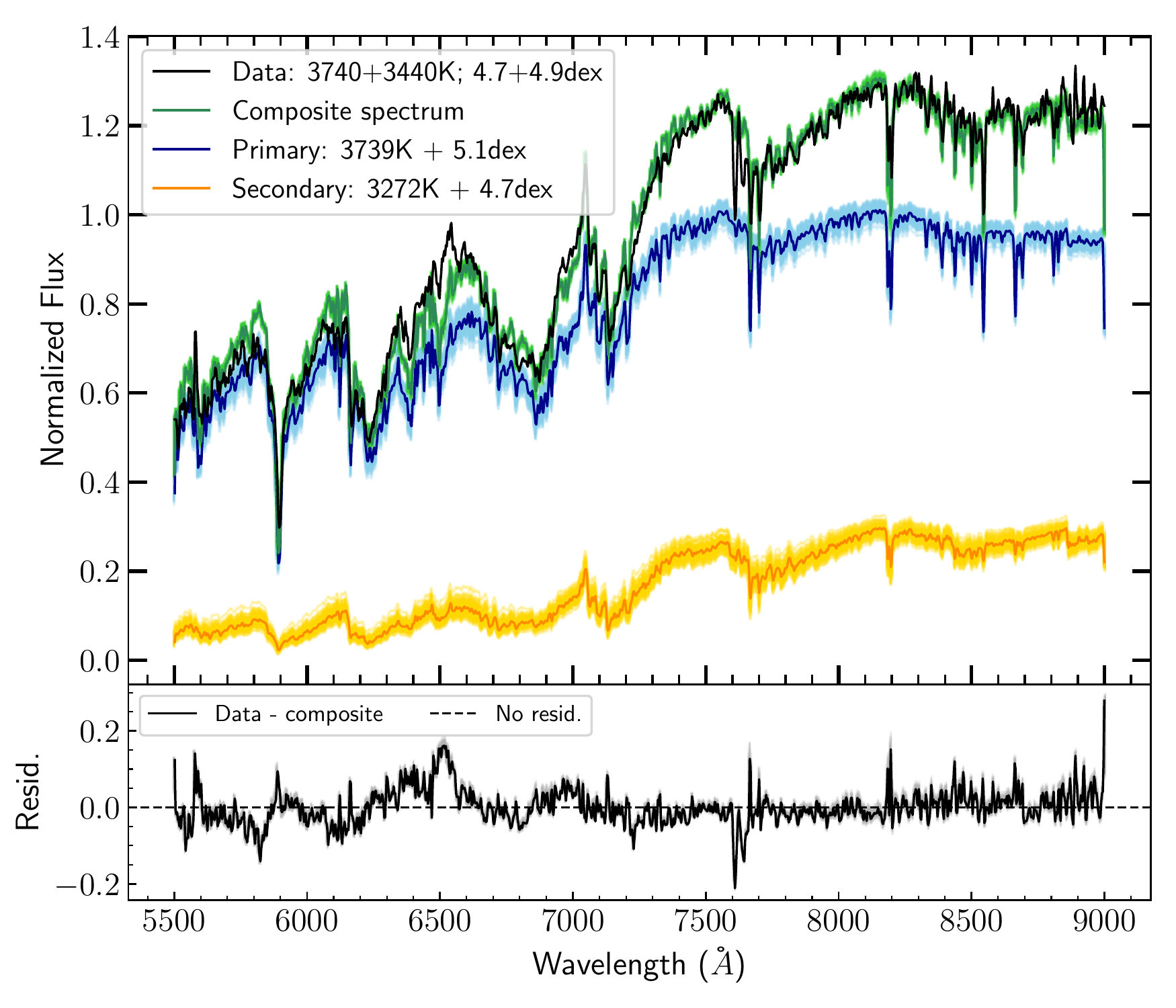} 
	\includegraphics[width = 0.45\linewidth]{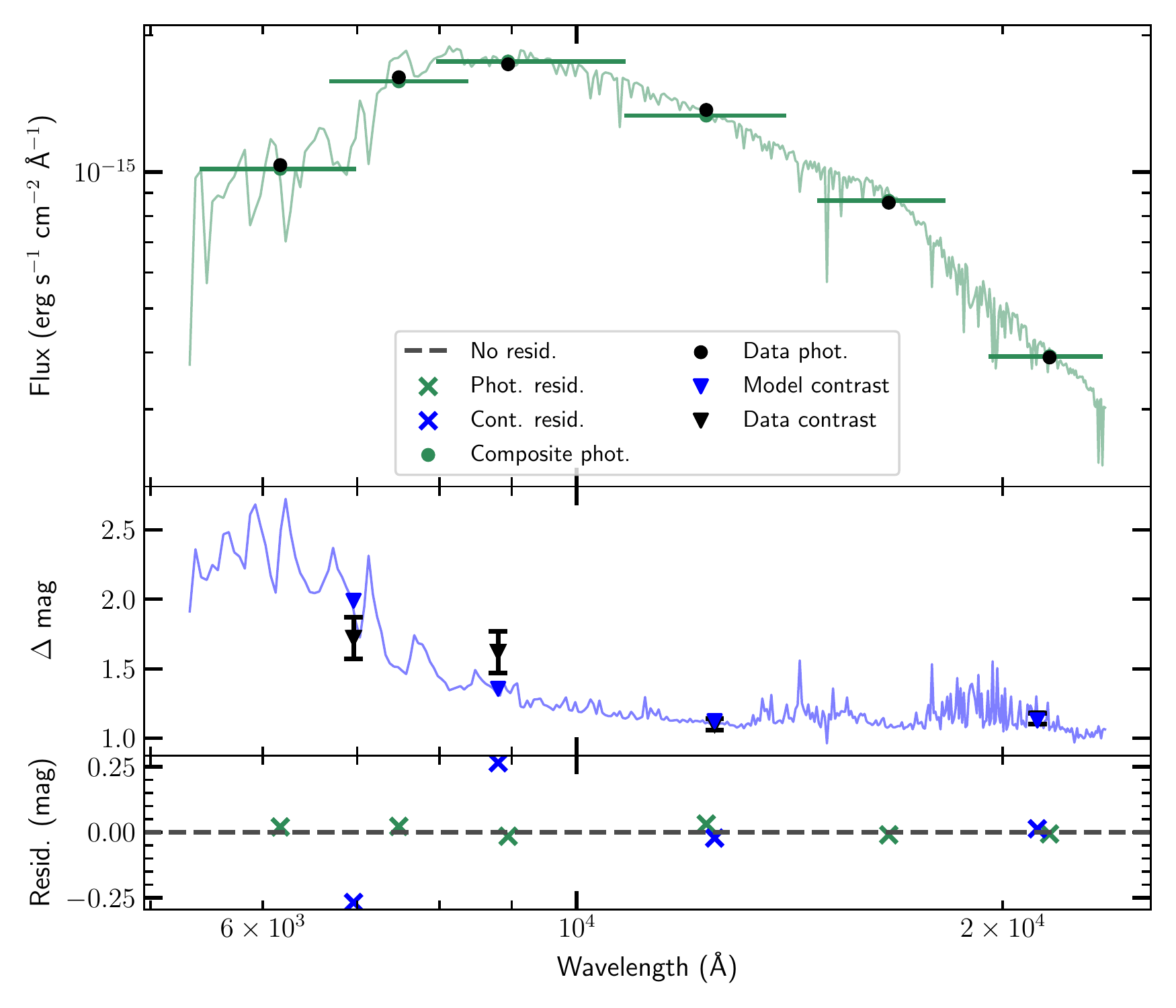}
}
\caption{Fit results for KOI-1422. Left: The best-fit spectrum (green), data (black) and primary and secondary component spectra (blue and gold, respectively) for KOI-1422. 100 random samples from the MCMC chain are also plotted for the three model spectra. The bottom panel shows the residuals between the data and the best-fit (black) and 100 random draws (gray) spectra. The legend shows the component temperatures and surface gravities. The broad spreads in the component spectra do not result in significant variance in the composite spectrum because of the covariance between primary and secondary star temperature. The residuals are typically small, except in specific poorly-modeled lines, which show mismatch on the order of 20\%. Right: The measured and best-fit model contrasts for KOI-1422. The top panel of the figure shows the observed KIC and 2MASS photometry (black points) and the best-fit model photometry (green points). The middle panel of the figure shows the observed(black triangle) and best-fit model (blue triangle) contrasts, while the bottom panel shows the residuals for the photometry (green x) and contrasts (blue x) in terms of $\Delta$ mag. All residuals are typically less than 0.25 mag.}
\label{fig:1422 spec}
\end{figure*}

\begin{figure*}
\plotone{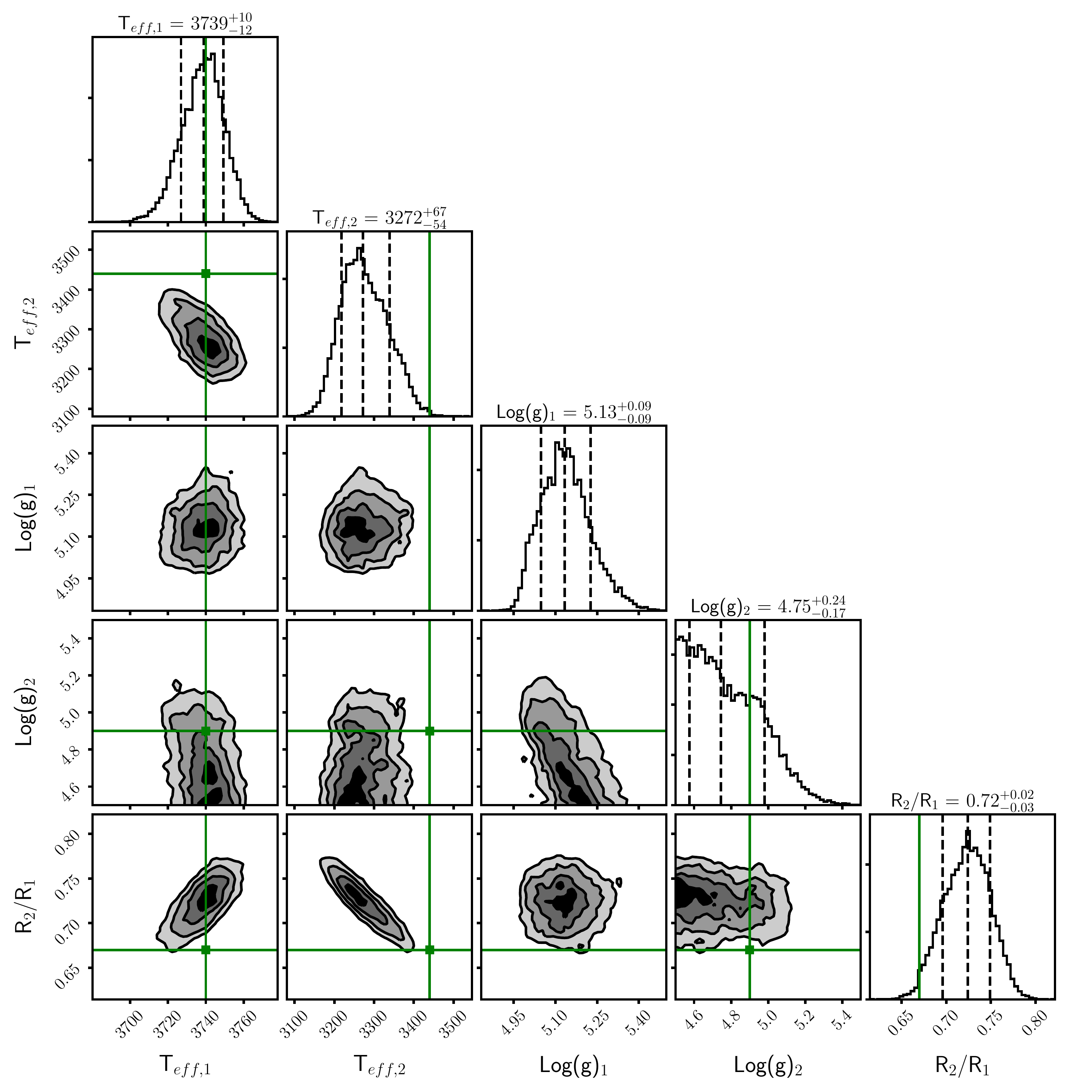}
\caption{A corner plot for KOI-1422 showing the marginalized posterior distributions and the covariance between different fit parameters. The green lines mark the values that were previously measured by B15. The secondary temperature, radii, extinction, and distance all match previously measured values well, but the secondary temperature is slightly higher than the previous measurement. The log(g) is poorly constrained for both components.}
\label{fig:1422 corner}
\end{figure*}

\begin{deluxetable*}{cccccc}
\tablecaption{Fit results for our method compared to B15 for KOI-1422\label{tab:koi1422}}
\tablecolumns{6}
\tablewidth{0pt}
\tablehead{
\colhead{Source} &
\colhead{T1} &
\colhead{T2} & \colhead{log(g)1} & \colhead{log(g)2}
& \colhead{R2/R1}\\
\colhead{} & \colhead{(K)} &
\colhead{(K)} & \colhead{(dex)} & \colhead{(dex)} & \colhead{}
}
\startdata
B15 & \val{3740}{130}{130} & \val{3440}{75}{75} & \val{4.77}{0.09}{0.06} & \val{4.93}{0.09}{0.06} & \val{0.67}{0.09}{0.11}\\
This work & \val{3739}{10}{12} & \val{3264}{66}{52} & $5.13 \pm 0.09$ & \val{4.75}{0.24}{0.17} & \val{0.72}{0.02}{0.03}\\
\enddata
\end{deluxetable*}

Using the two NIR contrasts (\textit{J} and $K_{S}$) from B15, the optical contrasts in $\Delta m_{[692]}$ and $\Delta m_{[880]}$ from \citet{Furlan2017}, and the M13 SNIFS-red spectrum (5100$< \lambda <$9000\AA), we deconvolved the stellar components. The best-fit model composite spectrum is shown with the data spectrum in Figure \ref{fig:1422 spec}, and the posterior distributions and covariance plots are shown in Figure \ref{fig:1422 corner}. We measured values of \teff$_{1}$ = \val{3739}{10}{12} K, \teff$_{2}$ = \val{3264}{66}{52} K, log(g)$_{1}$= $5.13 \pm 0.09$, log(g)$_{2}$ = \val{4.75}{0.24}{0.17}, and R$_{2}$/R$_{1}$ = \val{0.72}{0.02}{0.03}. Our results and the B15 results are listed in Table \ref{tab:koi1422}. The primary star temperature we measured for KOI-1422 was consistent with the B15 results, and the secondary star temperature was discrepant with B15 by $< 2 \sigma$ ($\sigma = 100$K). We measured smaller statistical errors than the errors in B15, and unlike their analysis we measured larger statistical errors in the temperature of the secondary star than in that of the primary star, because the secondary is the fainter component in the spectrum and so has larger intrinsic uncertainty. The measured radius ratio of KOI-1422 is consistent with the B15 value within the mutual uncertainties.

The surface gravity measured for the primary star is inconsistent with B15. Although the surface gravity of the secondary is technically consistent with the B15 measurement, the posterior is nearly flat, indicating that the surface gravity remains poorly constrained. These results show that our fitting method cannot recover surface gravity accurately in real data, confirming the conclusions drawn during the fit to synthetic data. This is likely because the spectral resolution is too low to accurately resolve the surface gravity-sensitive lines in the spectrum. We conclude that although we are able to accurately recover temperatures and radius ratio in real data, we are unable to accurately recover surface gravity.

\section{Results}\label{sec:new results}

\begin{deluxetable*}{lCCCCCCCCC} 
\tablecaption{Fit results for all KOIs \label{tab:star params}}
\tablecolumns{10}
\tablewidth{0pt}
\tablehead{
\colhead{KOI} &
\colhead{\teff$_{,1}$} &
\colhead{\teff$_{,2}$} & \colhead{R$_{1}$} & \colhead{R$_{2}$/R$_{1}$} & \colhead{\teff$_{,M13}$} & \colhead{\teff$_{,Kep}$} & \colhead{$\Delta m_{Kep}$} & \colhead{$f_{p, corr}$} & \colhead{$f_{s, corr}$}\\
\colhead{} & \colhead{(K)} &
\colhead{(K)} & \colhead{(\rsun)} & \colhead{} & \colhead{(K)} & \colhead{(K)} & \colhead{(mag)} & \colhead{} & \colhead{}
}
\startdata
227 & 4475$^{+16}_{-21}$ & 3711$^{+31}_{-18}$ & \nodata & 1.24$^{+0.03}_{-0.02}$ & 4093 & 3745 & 0.83$^{+0.04}_{-0.05}$ & 1.21$^{+0.01}_{-0.01}$ & 2.23$^{+0.03}_{-0.01}$\\
1422 & 3739 $^{+10}_{-12}$ & 3274$^{+67}_{-54}$ & \nodata & 0.72$^{+0.02}_{-0.03}$ & 3622 & 3526 & 1.86$^{+0.07}_{-0.15}$ & 1.082$^{+0.016}_{-0.001}$ & 1.84$^{+0.12}_{-0.15}$\\
1681 & 3875$^{+31}_{-31}$ & 3488$^{+46}_{-46}$ & \nodata & 1.04$^{+0.07}_{-0.07}$ & 3700 & 3638 & 0.72$^{+0.19}_{-0.19}$ & 1.22$^{+0.02}_{-0.02}$ & 1.90$^{+0.07}_{-0.07}$\\
2124* & 4450$^{+33}_{-33}$ & 3905$^{+41}_{-41}$ & 0.56 $^{+0.02}_{-0.02}$ & 1.14 $^{+0.05}_{-0.05}$ & 4132 & 4132 & 0.47$^{+0.10}_{-0.17}$ & 1.270$^{+0.057}_{-0.008}$ & 1.92$^{+0.04}_{-0.29}$ \\
2174 & 4379$^{+43}_{-97}$ & 4129$^{+98}_{-58}$ & \nodata & 1.03$^{+0.04}_{-0.06}$ & 4246 & 4245 & 0.32$^{+0.06}_{-0.14}$ & 1.32$^{+0.04}_{-0.012}$ & 1.61$^{+0.05}_{-0.18}$\\
2542 & 3506$^{+28}_{-35}$ & 3371$^{+92}_{-98}$ & \nodata & 0.69$^{+0.04}_{-0.04}$ & 3443 & 3460 & 1.15$^{+0.21}_{-0.19}$ & 1.16$^{+0.03}_{-0.02}$ & 1.37$^{+0.18}_{-0.17}$\\
2862 & 3900$^{+47}_{-41}$ & 3688$^{+47}_{-38}$ & \nodata & 1.08$^{+0.04}_{-0.04}$ & 3800 & 3678 & 0.202$^{+0.137}_{-0.018}$ & 1.341$^{+0.017}_{-0.025}$ & 1.68$^{+0.03}_{-0.14}$\\
3010 & 4197$^{+55}_{-50}$ & 3718$^{+66}_{-64}$ & \nodata & 1.05$^{+0.06}_{-0.06}$ & 3955 & 3808 & 0.74$^{+0.16}_{-0.06}$ & 1.206 $^{+0.032}_{-0.007}$ & 1.87$^{+0.12}_{-0.17}$\\
\enddata
\tablecomments{An asterisk (*) denotes a system where a Gaia parallax was available. If the R$_{1}$ column shows \nodata, it means that the system had no parallax so we did not fit for the absolute primary star radius, only the radius ratio. The Kepler temperature value is retrieved from the Kepler team data re-analysis presented in \citet{Furlan2017} and listed on the ExoFOP. The Kepler and M13 temperatures are the values that were measured for the unresolved binaries in the KIC and M13. $\Delta m_{Kep}$ was calculated using the spectra and radius ratio from a random sample of 1000 entries in the MCMC chains. $f_{corr}$ is the planet radius correction factor if either the primary or secondary star is the planet host, derived using the method introduced in \citet{Ciardi2015} and the inferred distribution of \kep\ contrasts from the MCMC analysis.}
\end{deluxetable*}

After validating our method with several different test systems, we fit 7 previously unresolved systems with unresolved spectra from M13 and contrasts at multiple wavelengths. The best-fit measured primary and secondary star temperatures, radius ratio, and primary radius (if the system has a \textit{Gaia} parallax) are listed in Table \ref{tab:star params}, along with the previously measured temperatures from the ExoFOP and M13. The appendix (Appendix \ref{sec:appendix}) shows summary plots of the fits for each system included in our analysis formatted the same was as our results for KOI-1422 (Figures \ref{fig:1422 spec} and \ref{fig:1422 corner}). There were two systems, KOIs 1681 and 2124, that exhibited bimodal posteriors for temperatures and radius ratio. For those systems we imposed a weak prior by fitting a double-peaked Gaussian distribution to the posterior, then assumed that the mean and standard deviation of the Gaussian with the largest area were the most probable underlying values for the temperatures and radius ratio. In both cases, the chosen mode was more consistent with having two stars drawn from the same isochrone in the HR diagram and having masses that were roughly consistent with their temperatures \citep{Mann2019}.

Table \ref{tab:star params} presents the most likely contrast in the \kep\ bandpass, which we calculated by drawing component spectra and their corresponding radius ratios from a set of 1000 randomly selected \emcee\ samples and calculating the \kep\ magnitude for each component using the same methodology as our other contrast measurements (Section \ref{subsec:phot}). In the \kep\ bandpass, the majority of the systems have contrasts between $0.5 < \Delta m < 1$ mag. However, none of the systems had a large enough separation to be resolved by \kep. The relatively small average contrast between the two components implies that either component could have produced a detectable signal from a transiting planet, and therefore that regardless of host star the planet's radius must be substantially revised.

We also calculated the planet radius correction factor $f_{corr}$. The correction factor if the primary star is the planet host is $f_{corr, pri}$, while $f_{corr, sec}$ is the correction factor if the secondary star is the planet host. $f_{corr}$ is defined such that R$_{p, corr} = f_{corr}$R$_{p, obs}$, and $f_{corr, pri} = \sqrt{1 + 10^{-0.4\Delta m}}$, while $f_{corr, sec} = \frac{R_{2}}{R_{1}}\sqrt{1 + 10^{+0.4 \Delta m}}$ \citep{Ciardi2015, Furlan2017}. The radii of planets around the primary stars would be revised upward by 23\% on average, while the radii of planets around the secondary stars would be revised upward by an average of 80\%.

\begin{figure}
    \plotone{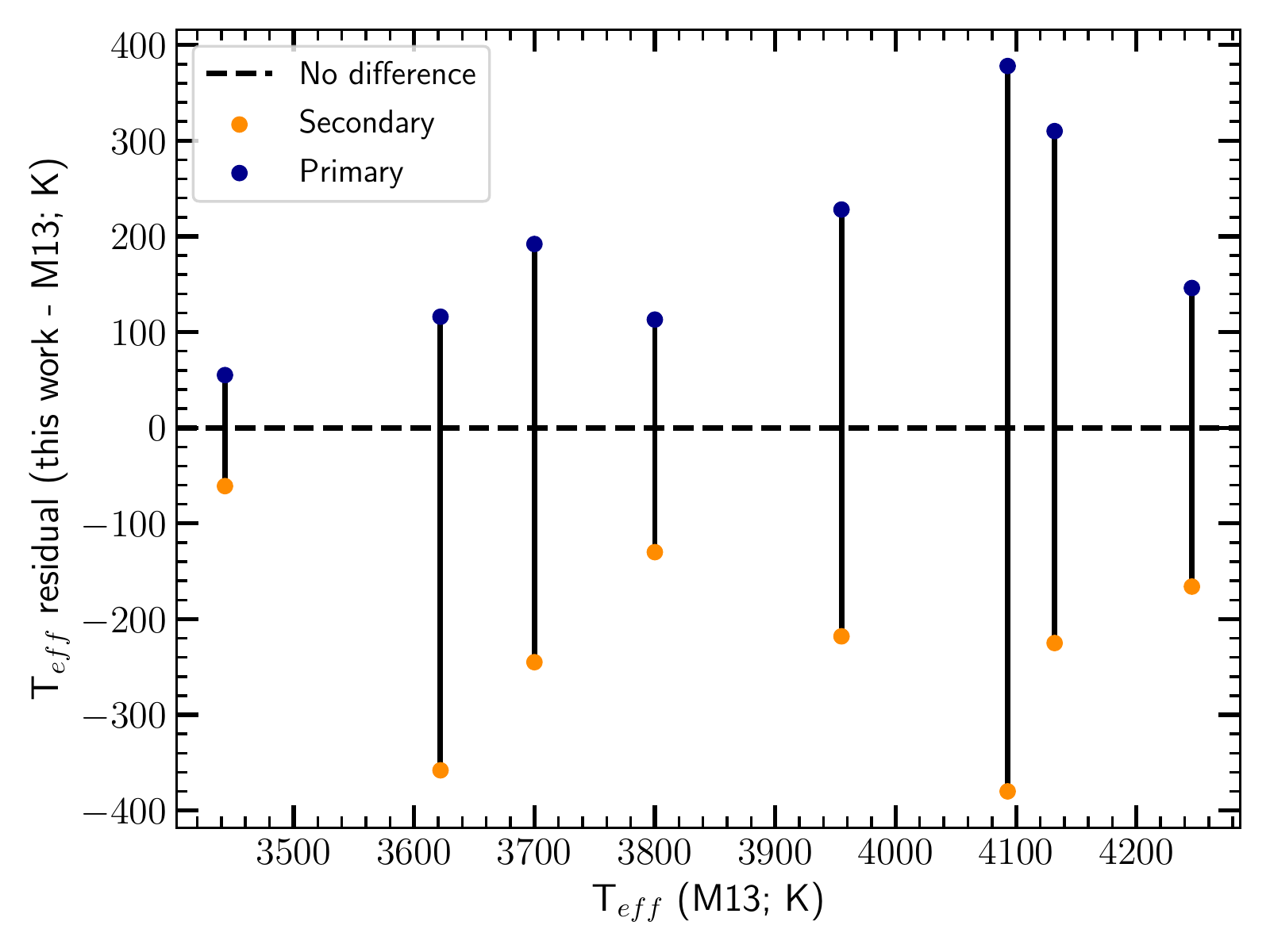}
    \caption{The difference between the primary and secondary temperatures and the measured M13 temperature, plotted as a function of measured M13 temperature. The black lines between the primary and secondary star points connect each system.}
    \label{fig:teff_diff}
\end{figure}

The temperatures of the primary stars in our sample were revised upward by an average of +192$\pm$101 K (minimum = +55 K; maximum = +378 K). Figure \ref{fig:teff_diff} visualizes the temperature difference for each system for both the primary and secondary stars relative to the M13 measured temperature. Although we did not restrict the primary star (i.e., the star that the contrast was measured in reference to) to be the hotter star in the system, the brighter star was always measured to be hotter. In all cases the retrieved temperatures for the two component stars bracketed the measured M13 temperature for the composite system, which was consistent with our expectation that two stars should combine to create a composite temperature that falls between the true component temperatures. 

We measured radius ratios that were larger than unity for almost all systems. We did not observe this effect in any of our test cases, including KOI-1422, the analysis that used identical data to the other objects presented in this work. The magnitude of the effect does not appear to be dependent on the number of contrasts used in the fit. There are no apparent systematic trends for individual filters causing the anomalous radius ratio measurements. Instead, we note that many of these systems have $\Delta$K measurements that are close to zero, which can be verified through inspection of archival Keck Observatory Archive images. However, the systems also have nonzero optical contrasts, meaning that the secondary contribution must be near equal to the primary in K band and fainter than the primary in the optical. The only way to achieve this set of parameters is a secondary that is cooler and larger than the primary. The two systems with the largest K-band contrasts, KOI-1422 and KOI-2542, both have radius ratios $R_{2}/R_{1} < 1$, supporting the possibility that the contrasts are driving the radius ratio results.

Another possibility for some systems having radius ratio measurements greater than unity is the presence of triple stars in our sample. For example, KOI-227, with a measured radius ratio of $R_{2}/R_{1} = 1.23$, has the largest measured radius ratio and also has contrasts that are poorly fit by a two-star model. However, only $\sim 8\%$ of stars are triples \citep{Raghavan2010} so in the sample of 8 stars it is statistically unlikely that there are more than one or two undetected triples.

\section{Discussion}\label{sec:disc}
To revise the properties of \kep\ binary planet host stars, we have constructed an algorithm that uses the BT-Settl stellar atmosphere models to fit low-resolution spectra of unresolved binary stars and retrieve the individual stellar parameters. We have tested our algorithm on a single star, a synthetic spectrum, and a previously characterized KOI, then applied our fitting process to 7 additional M-dwarf planet hosts with pre-existing data from M13. 

We found that accounting for multiplicity and fitting for component spectra means that derived component temperatures can be significantly different from the composite measured system temperature. In this section we discuss the scale of the temperature changes accounting for multiplicity causes (Section \ref{subsec:impact}) and assess the quality of our derived parameters (Section \ref{subsec:quality}). Section \ref{subsec:error} discusses the statistical versus systematic error on our fits.

\subsection{The Impact of Neglecting Multiplicity}\label{subsec:impact}
Figure \ref{fig:teff_diff} shows that are no systematic trends between composite system \teff\ and resolved primary and secondary \teff, emphasizing that many different combinations of primary and secondary \teff\ can result in similar measured unresolved \teff. Thus, establishing the true primary star \teff\ for unresolved binary stars via multi-component fitting is particularly important, because there is no systematic correction that can be made for multiplicity: A measured unresolved \teff\ is not predictive of a particular primary or secondary star \teff. 

\citet{Furlan2020} conducted an analysis that was functionally the inverse of this work by taking high-resolution California-Kepler Survey (CKS; \citealt{Johnson2017, Petigura2017}) spectra of single stars with different temperatures and combining them with various radial velocities. They then characterized the composite spectra to constrain the temperature bias and other systematic biases, like surface gravity and radius, induced by the presence of secondary stars at given temperature (and analogously, luminosity) differentials. \citet{Furlan2020} found that the measured temperature changes caused by the presence of a secondary were rarely larger than $\sim$ 200 K, and the most common temperature shift resulting from a binary was on the order of 50 K.

We found that the average temperature difference between composite temperatures and primary star temperatures for M star binaries was $\sim$ 200 K rather than the typical 50 K found by \citet{Furlan2020}, meaning that the effects of multiplicity on temperature measurements of M stars is more pronounced than in the \citet{Furlan2020} sample of G and K stars. This may be because even at low resolution, M star optical spectra are dominated by molecular TiO bands that are temperature sensitive, meaning that even a relatively cool (and thus faint) secondary star can still significantly alter the measured temperature. Additionally, spectra of hotter stars have fewer large spectral features, meaning that their contribution to a spectrum may be masked by the large, information-rich features typical of M star spectra. Because low-mass secondary stars are common, this contamination may be a concern for a wider range of temperatures and mass ratios than were explored in the current study, which was restricted to M + M and K + M binaries. Future work will examine samples of binary stars with a wider range of stellar temperatures for both primary and secondary stars, and may be able to better constrain the effects of different secondary star spectral types on temperature measurements.

\subsection{Quality of Retrieved Parameters}\label{subsec:quality}
All of the systems except for KOI-1422 and KOI-2542 had radius ratios measured to be greater than unity. The origin of this discrepancy is unclear, and has several possible explanations. One cause of large radius ratios could the the presence of a third cool component in the system, which would cause a larger contribution from a cool component in the spectrum and thus result in a radius ratio greater than unity. This may be the case for KOI-227, which has the largest measured radius ratio in our sample and also has contrasts that are not well-fit by a two-star model. In general, our sample should not be exclusively composed of triple stars, and other options are more likely, especially because the majority of our systems have contrasts that are well-fit by the two-star model.

Another possible factor impacting the measured radius ratios is the precision of the available contrasts and photometry, which may not be adequate to constrain the contrast between the two stellar components. We tested running fits with artificially reduced error bars for both contrasts and photometry separately to explore the impact more tightly constrained photometry would have on our results. We found that reducing the unresolved photometry error bars did not significantly change the radius or temperature results. Artificially reducing the contrast errors typically reduced the measured radius ratio to less than unity and occasionally altered the measured temperatures by $\sim 100$K, indicating that more precise high-resolution imaging would produce more accurate radius ratio measurements and might slightly change the derived temperatures. 

One resolution to measuring radius ratios that are greater than unity and appear to be nonphysical would have been to impose a prior on the stellar radii by forcing the two components of the binary to fall on the same isochrone. However, we expected the binaries to be coeval, and thus to fall on the same isochrone naturally, so removing the prior of an isochrone from the systems was an independent check on the fitting method. The lack of a prior on the radius ratio was how we found that the optical and NIR contrasts conflicted and produced radius ratios larger than unity, which we would not have discovered in the presence of the prior.

We measured bimodal posterior distributions in temperature and radius ratio for two systems, KOI-1681 and KOI-2124. In the case of KOI-1681, we interpret the bimodality as being caused by some walkers in the \emcee\ run getting stuck in the ``inverted'' temperature regime where T1 $<$ T2. This regime is not strongly disallowed by the data because we had to reject one optical contrast (Section \ref{sec:sample}) and the other has a large uncertainty (0.15 mag; Table \ref{tab:star params}). In the case of KOI-2124, we interpret the bimodality as a result of the bimodal surface gravity posterior, which seems to be caused by the fitting algorithm seeking to compensate for the spectral features caused by multiplicity by altering spectral lines that are sensitive to surface gravity. In each case, we imposed a weak prior that T1 $>$ T2, which we implemented by fitting a two-component Gaussian distribution and choosing the parameters of the Gaussian that had the larger area. We chose not to impose the prior directly so that we could present results that were unconstrained by astrophysics, i.e., the same reason we did not impose a radius prior. For these two systems those are the values reported in Table \ref{tab:star params}, but the full posteriors, including the 16th, 50th, and 84th percentiles are reported in the corner plots in Appendix \ref{sec:appendix}. In the future, a weak prior on the surface gravity may be appropriate to prevent it from ranging into nonphysical regimes (e.g., $\log(g) = 5.5$ for a main sequence M dwarf).

\subsection{Assessing Systematic Error on Fits}\label{subsec:error}
The errors presented in Table \ref{tab:star params} are the statistical error derived using \emcee, and do not incorporate any estimate of the systematic error in the temperature and radius measurements. One of the key sources of systematic error in our results is the error intrinsic to the model spectra we used in our fitting procedure. M13 and M15 found $\sim 60$ K disagreement between temperatures derived for M stars using different model spectra, suggesting that a 60 K systematic error is a realistic lower limit for temperature measurements of single stars. We also found that the synthetic systems with the largest offset from their true values typically had deviations of $\sim$ 50 K, and also found that the systems with the largest offset correspondingly had larger statistical errors (Figure \ref{fig:synth summary}), suggesting that the statistical error scales with systematic error. 

Another way to assess the scale of the systematic error of our fits is to perform model-to-model comparisons of the same systems. To that end, we fit all our systems with another version of the BT-Settl/\citet{Caffau2011} models \footnote{\url{https://phoenix.ens-lyon.fr/Grids/BT-Settl/CIFIST2011c/SPECTRA}} and measured the difference in derived temperature between the two versions of the model grid. The median change in each temperature was $\Delta T_{1} = 29$K, $\Delta T_{2} = 10$K, suggesting that the systematic error, at least between two model grids using similar line lists and the same stellar atmospheres, is on the order of the statistical error. The 30 K error found during our investigation is smaller than the suggested 60 K systematic error from M13 and M15, and we would expect larger (but difficult-to-assess) systematic errors for binaries depending on the signal-to-noise ratio of the spectrum and the specific flux ratio between the components. Therefore, we suggest 60 K as a lower limit on our temperature systematic error but caution that it may be larger depending on each system's properties.

\section{Summary}\label{sec:summary}
We have developed an algorithm to retrieve statistically robust temperatures and radius ratios for spectroscopically unresolved binary stars using moderate-resolution spectroscopy, high-resolution imaging, and archival 2MASS and KIC unresolved photometry. We analyzed a sample of 8 binary M-dwarf Kepler Objects of Interest using archival spectroscopy and imaging, and derived temperatures and radius ratios for each component.

We found that the primary star temperatures are nearly 200 K hotter on average than temperature measurements of the composite systems, suggesting that fitting for individual stellar components should be an integral part of assessing the properties of planets in binary or higher-order multiple systems. The average planetary radius correction factors were $\sim 20$\% if the primary star were to be the planet host and $\sim80$\% if the secondary star were to be the planet host. The large scale of the radius correction factors means that regardless of which star is the planet host, the alteration of the planetary radius because of multiplicity is large. For the stars in this sample, it is difficult to argue for which star is the likely planet host in any of the binaries because the relatively small average contrast ($\sim$0.75 magnitudes) between the two components means that either binary component could host a detectable transiting exoplanet.
\\
\\

We thank the referee for their helpful comments and suggestions. K.S. acknowledges that this material is based upon work supported by the National Science Foundation Graduate Research Fellowship under Grant No. DGE-1610403. A.W.M. was supported through NASA’s Astrophysics Data Analysis Program (80NSSC19K0583). The authors acknowledge the Texas Advanced Computing Center (TACC) at The University of Texas at Austin for providing high-performance computing resources that have contributed to the research results reported within this paper. This publication makes use of data products from the Two Micron All Sky Survey, which is a joint project of the University of Massachusetts and the Infrared Processing and Analysis Center/California Institute of Technology, funded by the National Aeronautics and Space Administration and the National Science Foundation. This research has made use of the SVO Filter Profile Service (\url{http://svo2.cab.inta-csic.es/theory/fps/}) supported from the Spanish MINECO through grant AYA2017-84089. This research has made use of the VizieR catalogue access tool, CDS, Strasbourg, France (DOI : 10.26093/cds/vizier). The original description of the VizieR service was published in 2000, A\&AS 143, 23. This work has made use of data from the European Space Agency (ESA) mission {\it Gaia} (\url{https://www.cosmos.esa.int/gaia}), processed by the {\it Gaia} Data Processing and Analysis Consortium (DPAC, \url{https://www.cosmos.esa.int/web/gaia/dpac/consortium}). Funding for the DPAC has been provided by national institutions, in particular the institutions participating in the {\it Gaia} Multilateral Agreement. This research has made use of the Exoplanet Follow-up Observation Program website, which is operated by the California Institute of Technology, under contract with the National Aeronautics and Space Administration under the Exoplanet Exploration Program.

\software{
astropy \citep{astropy2013, astropy2018}, corner \citep{DFM2016}, emcee \citep{DFM2013}, matplotlib \citep{Hunter2007}, numpy \citep{Harris2020}, scipy \citep{Virtanen2020}, exoFOP \citep{exofop}
}

\appendix 
\section{Diagnostic plots for each system}\label{sec:appendix}
This appendix shows the diagnostic plots for each system analyzed in this work. On the top left of the figure, there is a plot showing the data spectrum, the best-fit composite spectrum, and the best-fit component spectra for the primary and secondary star. Underlaid on each model spectrum is a random sample of 100 spectra drawn from the MCMC samples to visually demonstrate the spread in the measurement. The bottom panel of the plot shows the residuals for each draw relative to the data, including the best-fit model in black and the 100 random spectra in gray. 

The top right plot in the figure shows the fit to the contrasts (middle panel) and the unresolved photometry (top panel), with a bottom panel showing the residuals to each fit. In all systems the residuals in the spectroscopy and contrasts are typically less than 0.1 mag. The best-fit spectrum and contrast curve are underlaid on the figures to aid in comparison between the best-fit parameters and the data.

The bottom row of the figure shows a corner plot with the results of the \emcee\ error determination. The diagonal plots show the marginalized posterior distributions for each parameter, while the off-diagonal plots show the covariance between different parameters. The vertical lines in the posteriors show the 16th and 84th percentiles, which are shown numerically as the errors in the measured parameters shown at the top of each column. For systems with a bimodal posterior, the values shown in the plot may not match the values in the text, because if a posterior was  bimodal we measured the best-fit parameters using a two-Gaussian fit to the posterior.

\begin{figure*}
\gridline{\includegraphics[width = 0.45\linewidth]{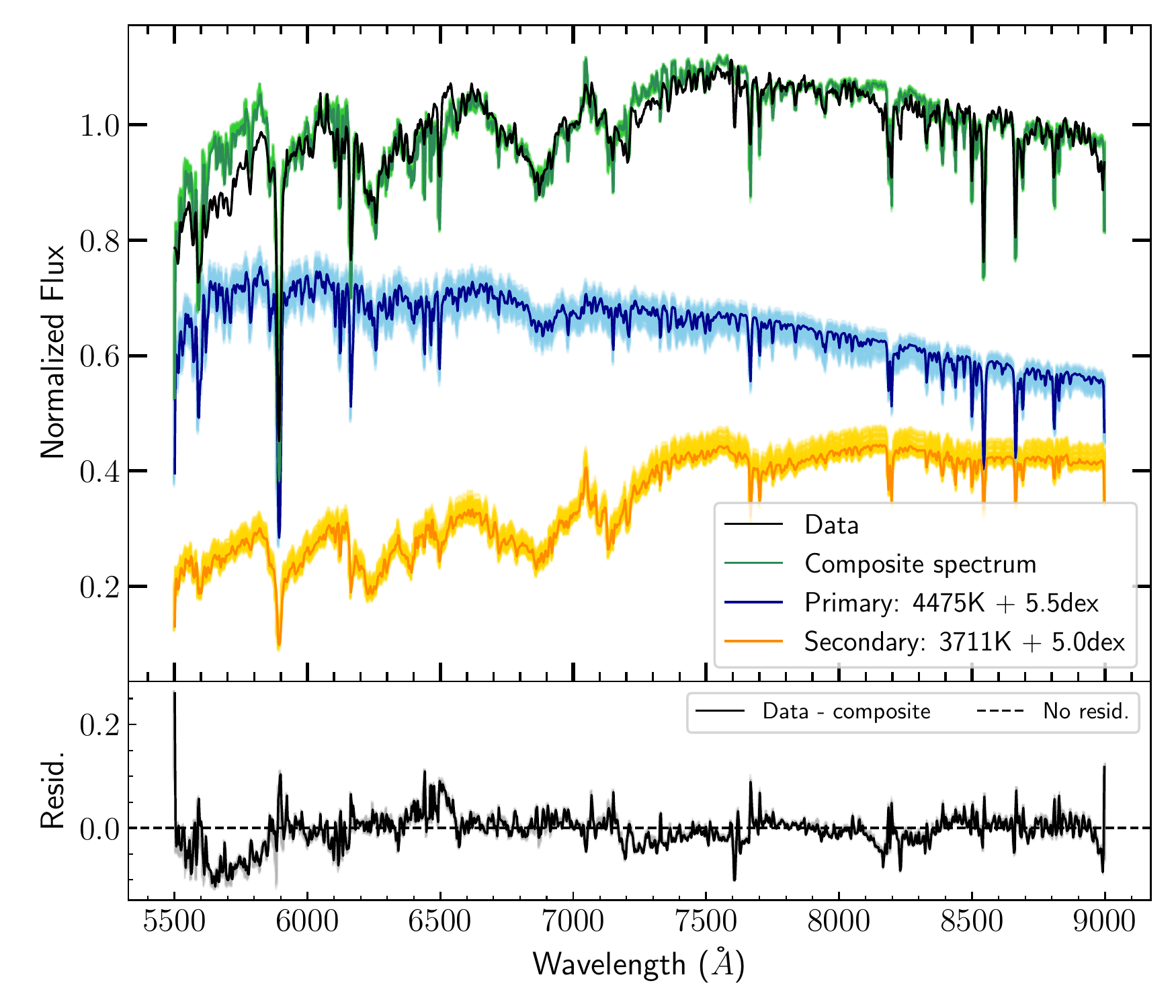} 
	\includegraphics[width = 0.45\linewidth]{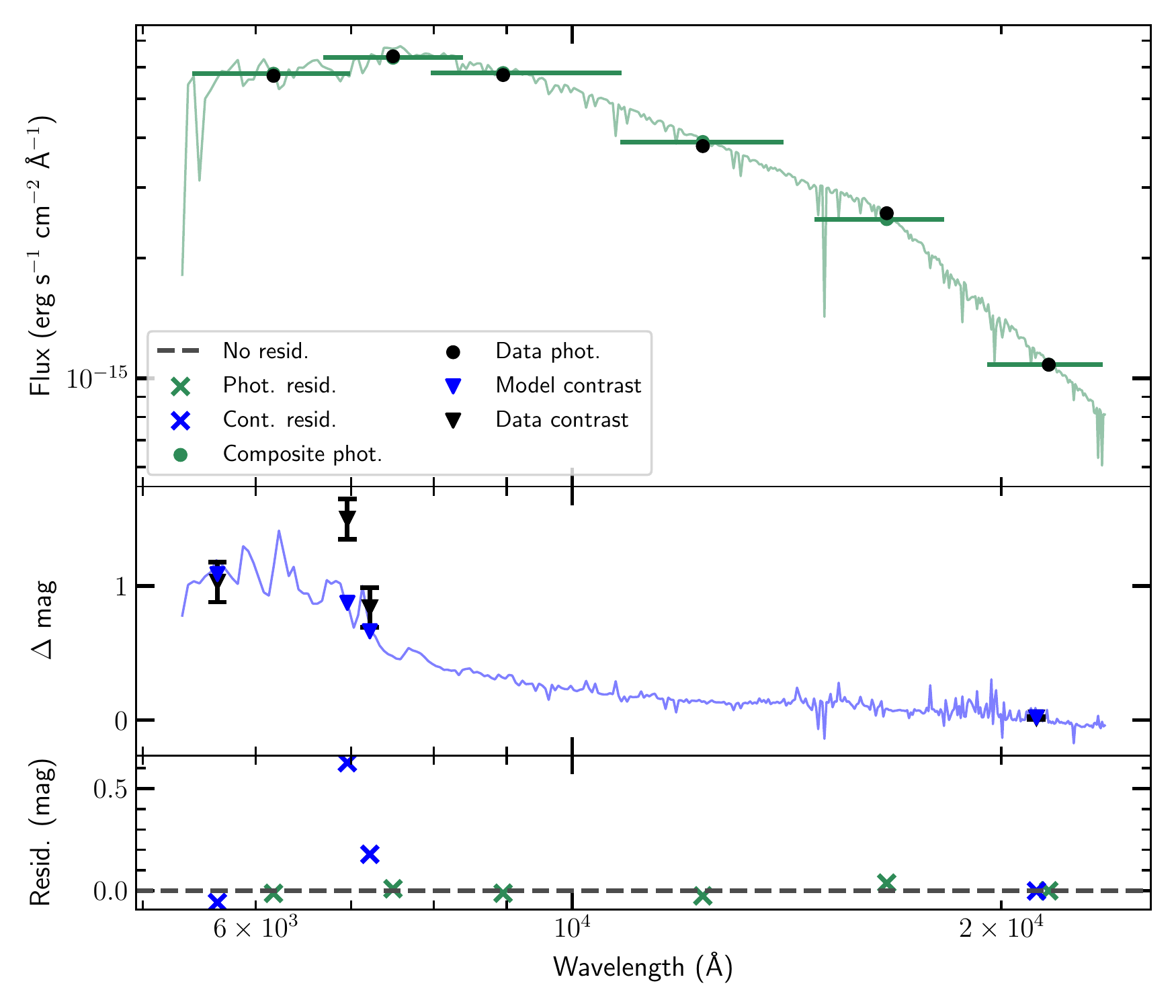}
}
\gridline{\includegraphics[width = 0.8\linewidth]{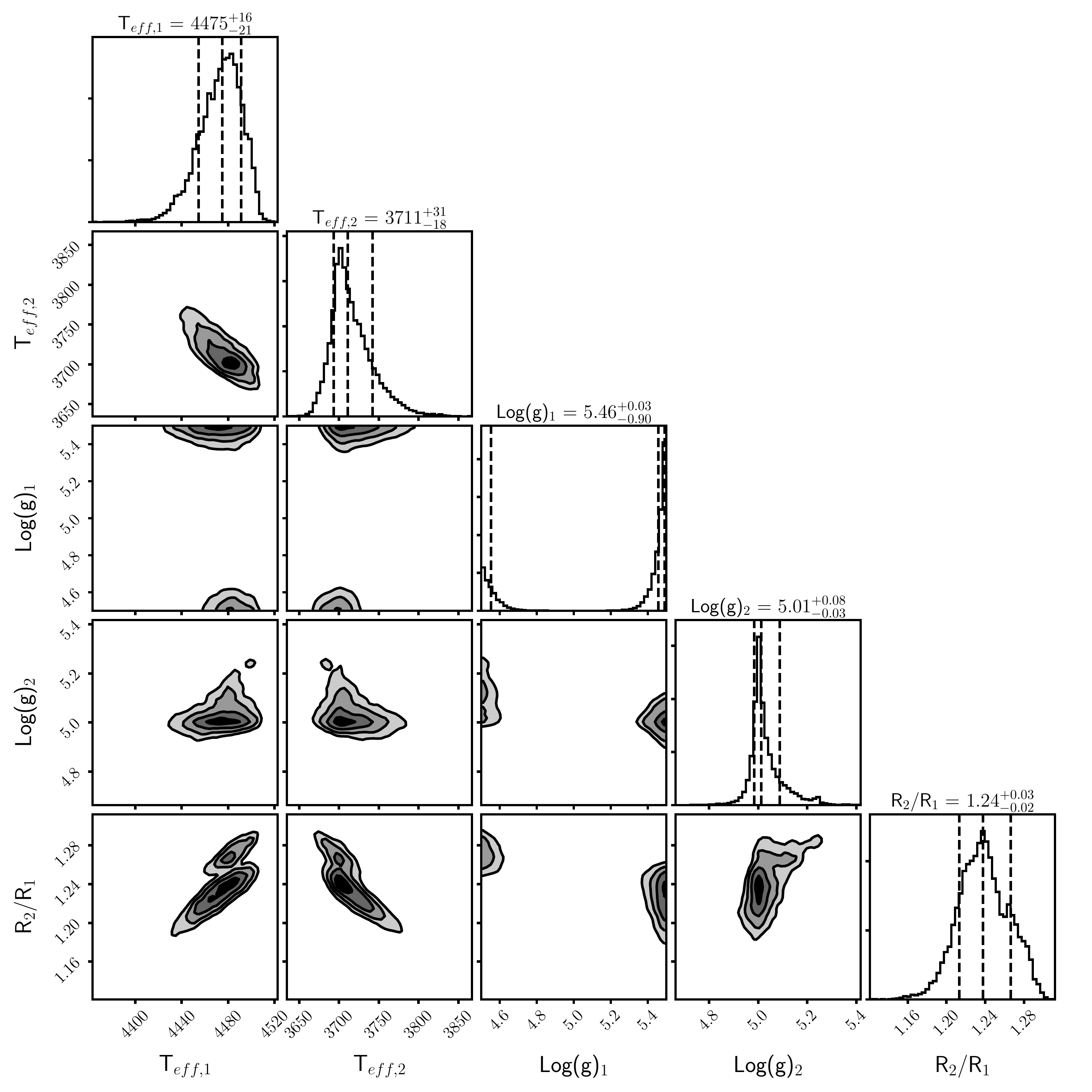}}
\caption{KOI-227}
\label{fig:227 spec}
\end{figure*}

\begin{figure*}
\gridline{\includegraphics[width = 0.45\linewidth]{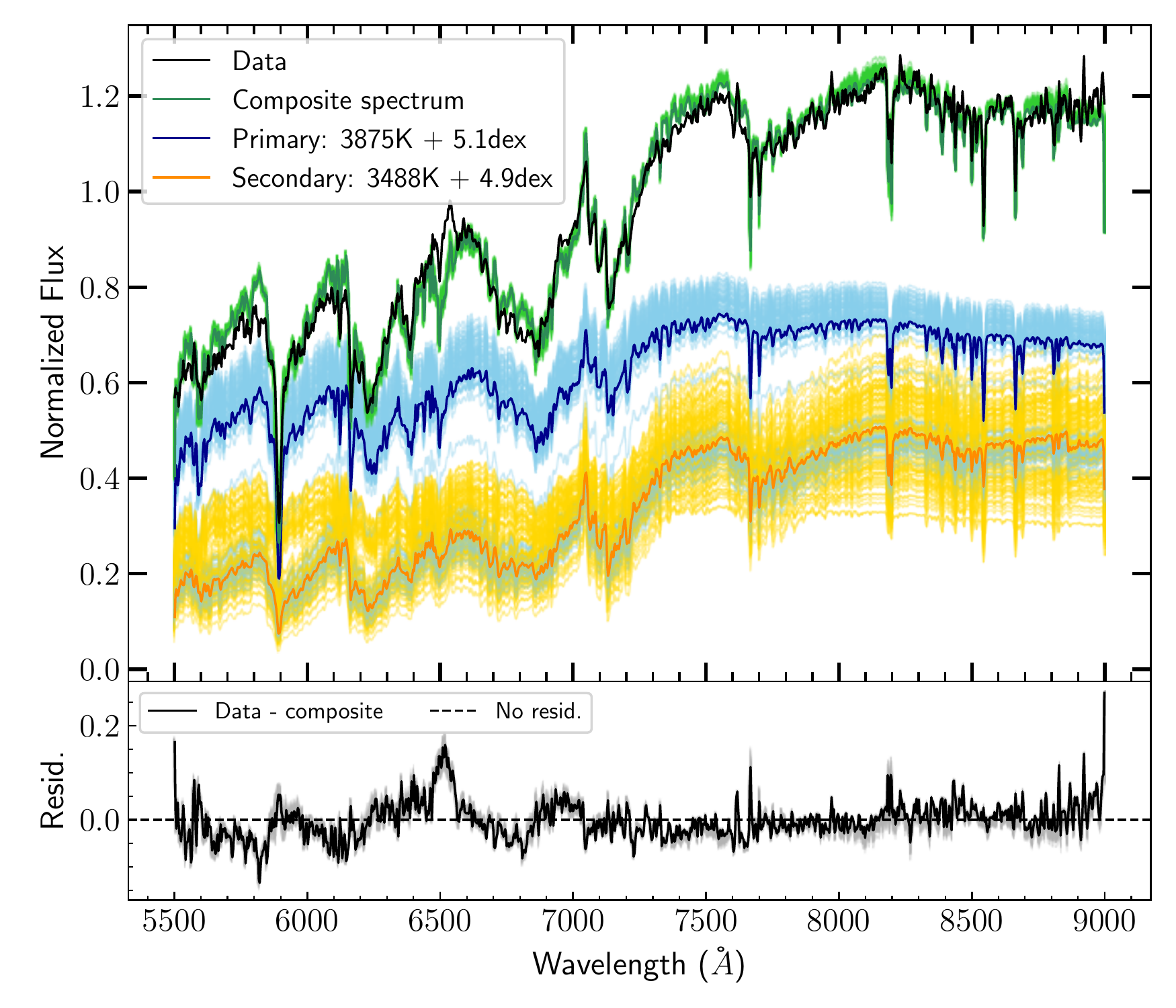} 
	\includegraphics[width = 0.45\linewidth]{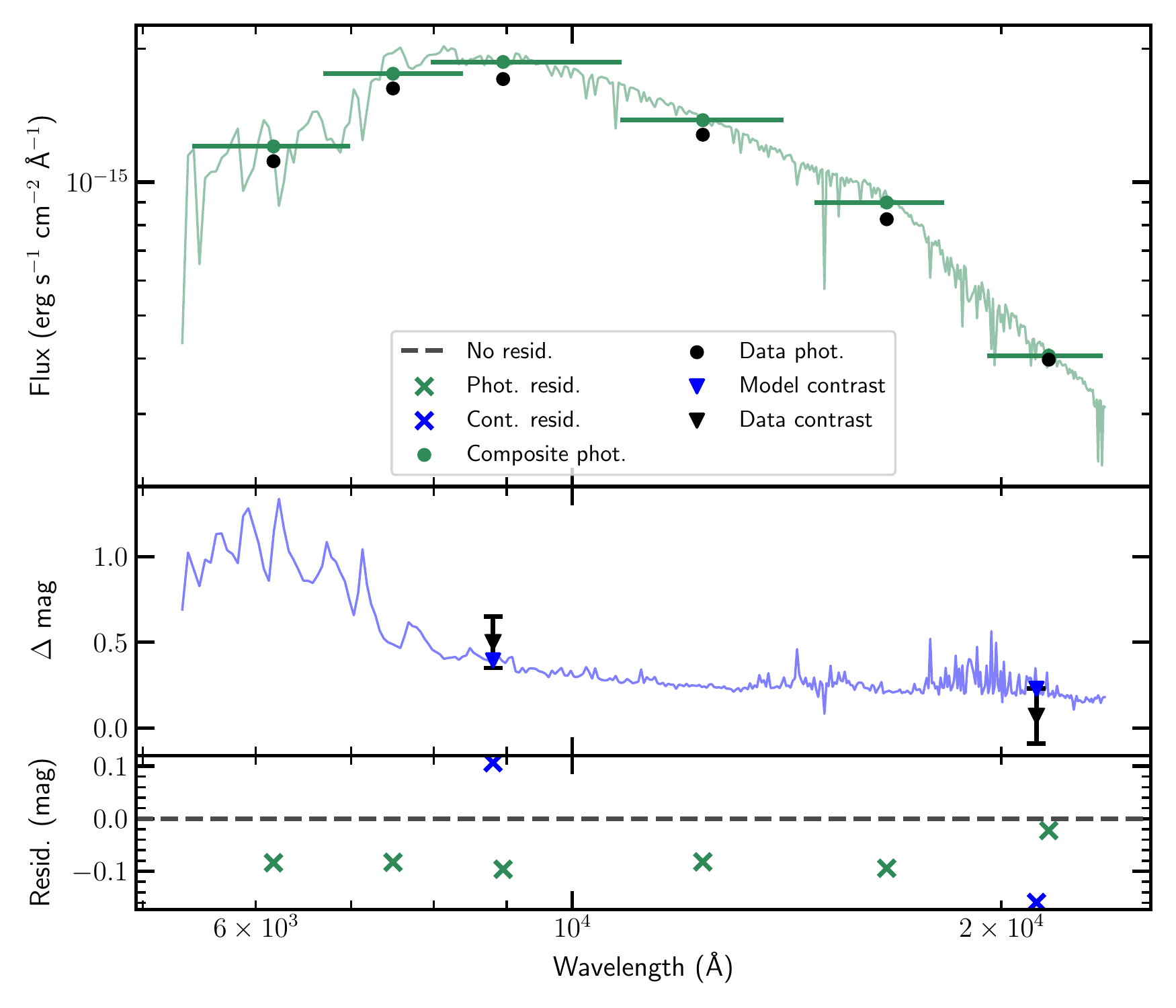}
}
\gridline{\includegraphics[width = 0.8\linewidth]{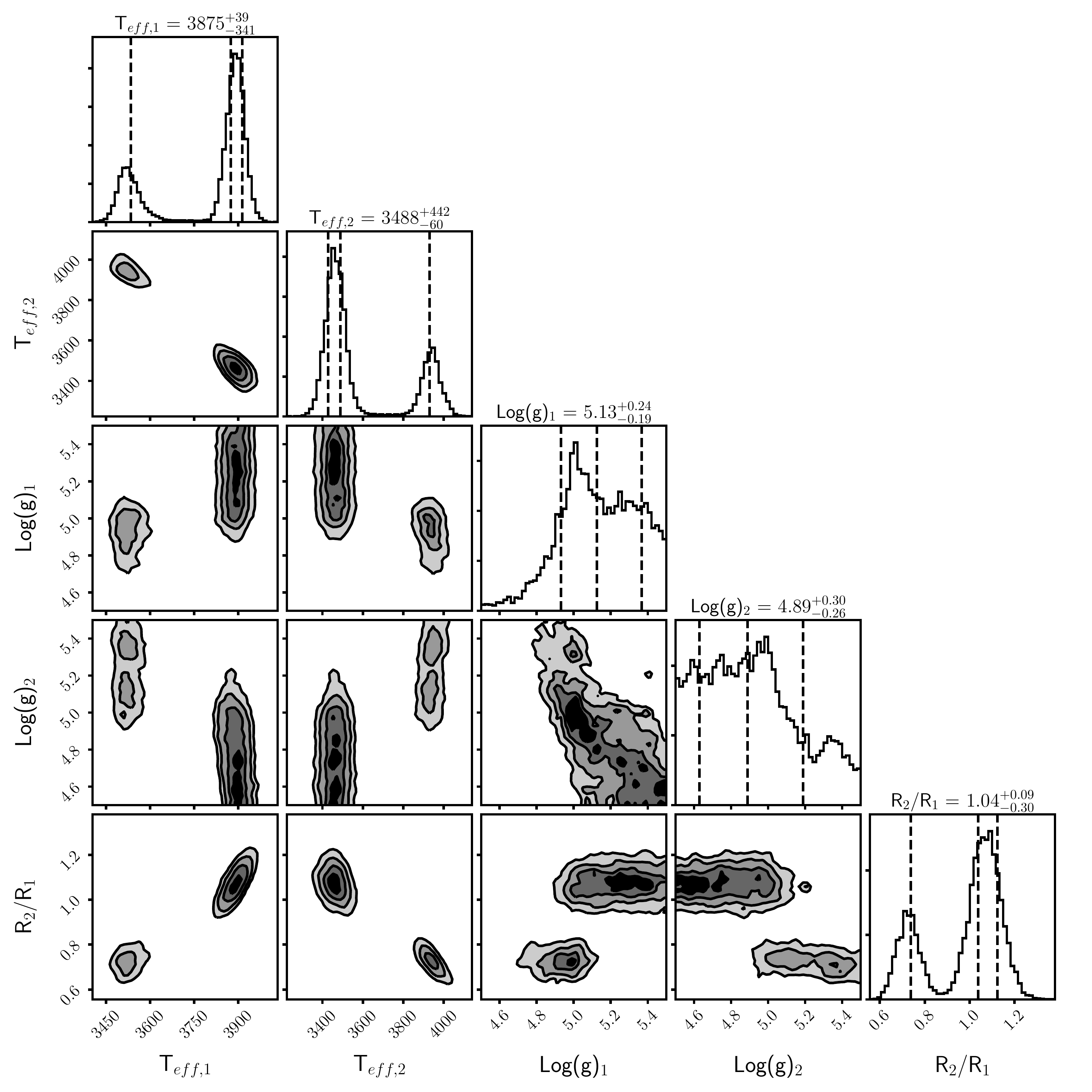}}
\caption{KOI-1681. We report only the dominant mode (that of T1 $>$ T2) in Table \ref{tab:star params}, but show the full posterior here for completeness.}
\label{fig:1681 spec}
\end{figure*}

\begin{figure*}
\gridline{\includegraphics[width = 0.45\linewidth]{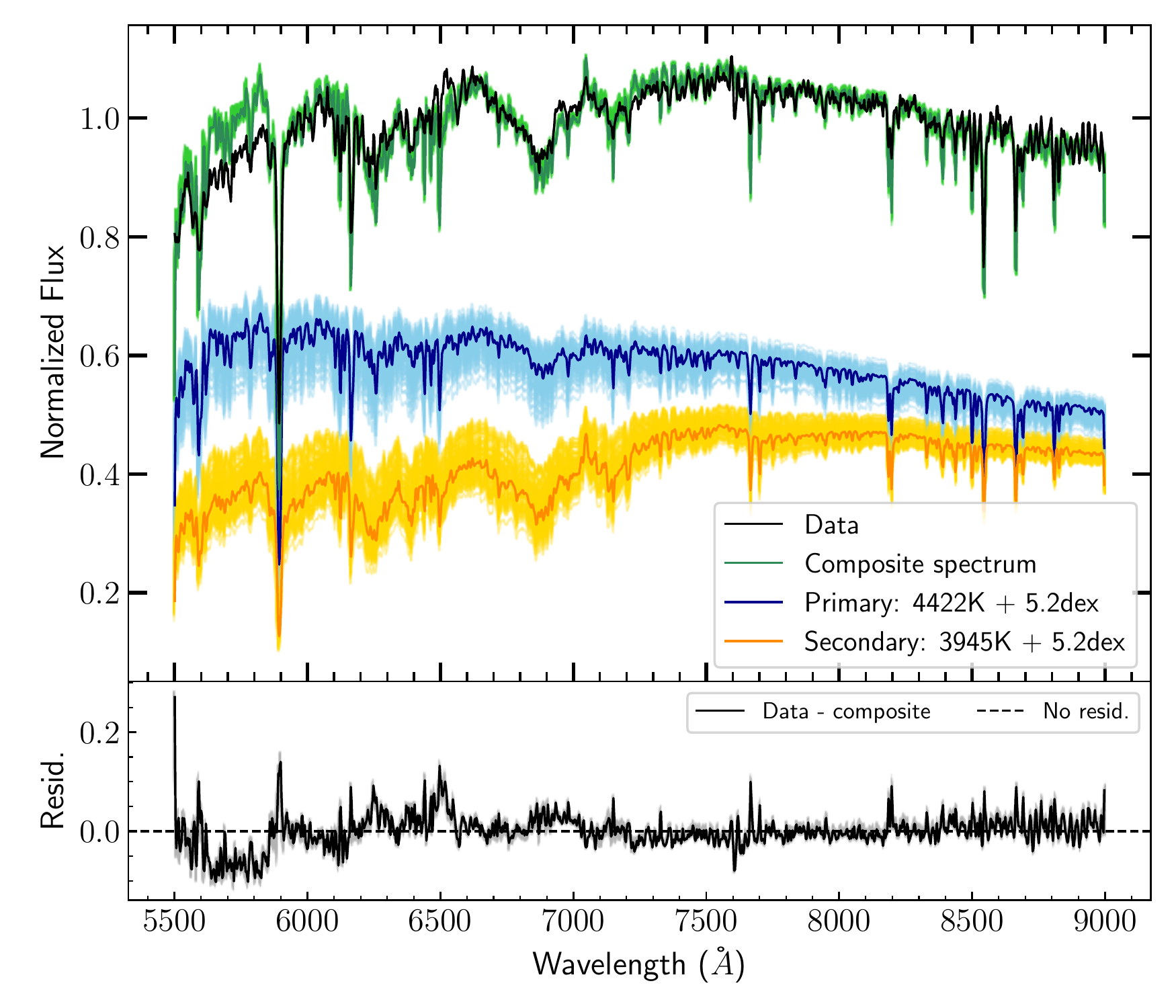} 
	\includegraphics[width = 0.45\linewidth]{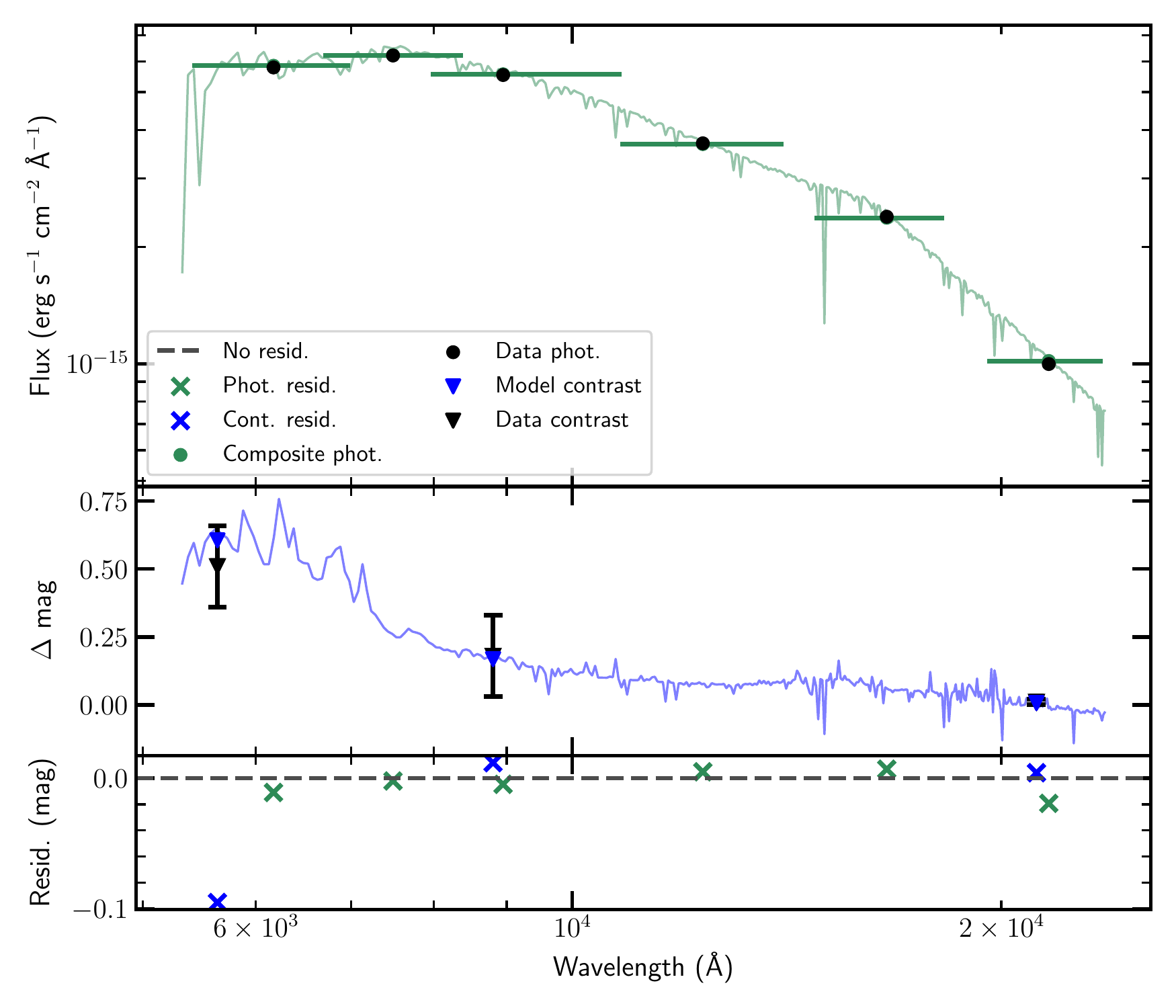}
}
\gridline{\includegraphics[width = 0.8\linewidth]{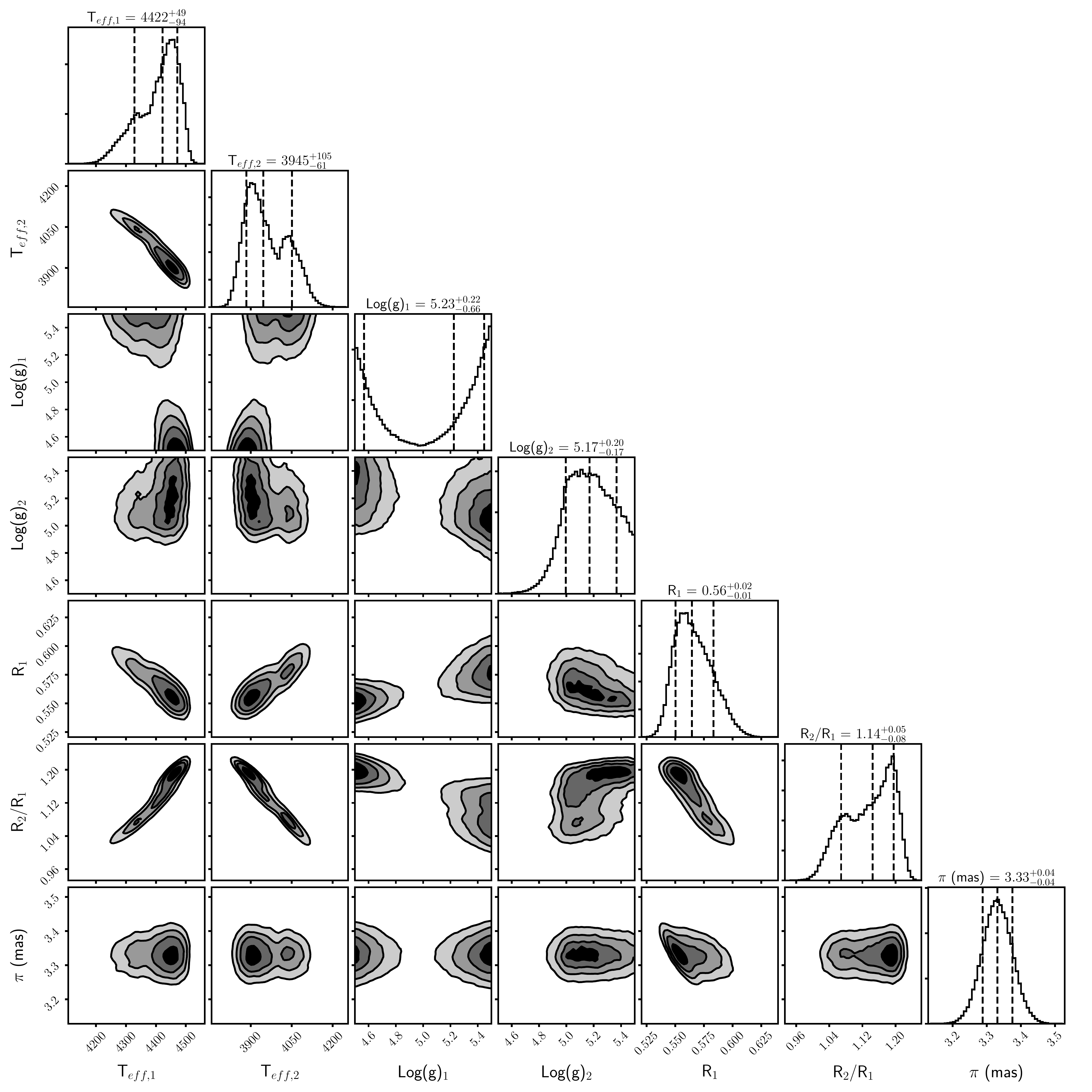}}
\caption{KOI 2124. We report only the dominant mode (that of log(g) $< 5.0$) in Table \ref{tab:star params}, but show the full posterior here for completeness.}
\label{fig:2124 spec}
\end{figure*}

\begin{figure*}
\gridline{\includegraphics[width = 0.45\linewidth]{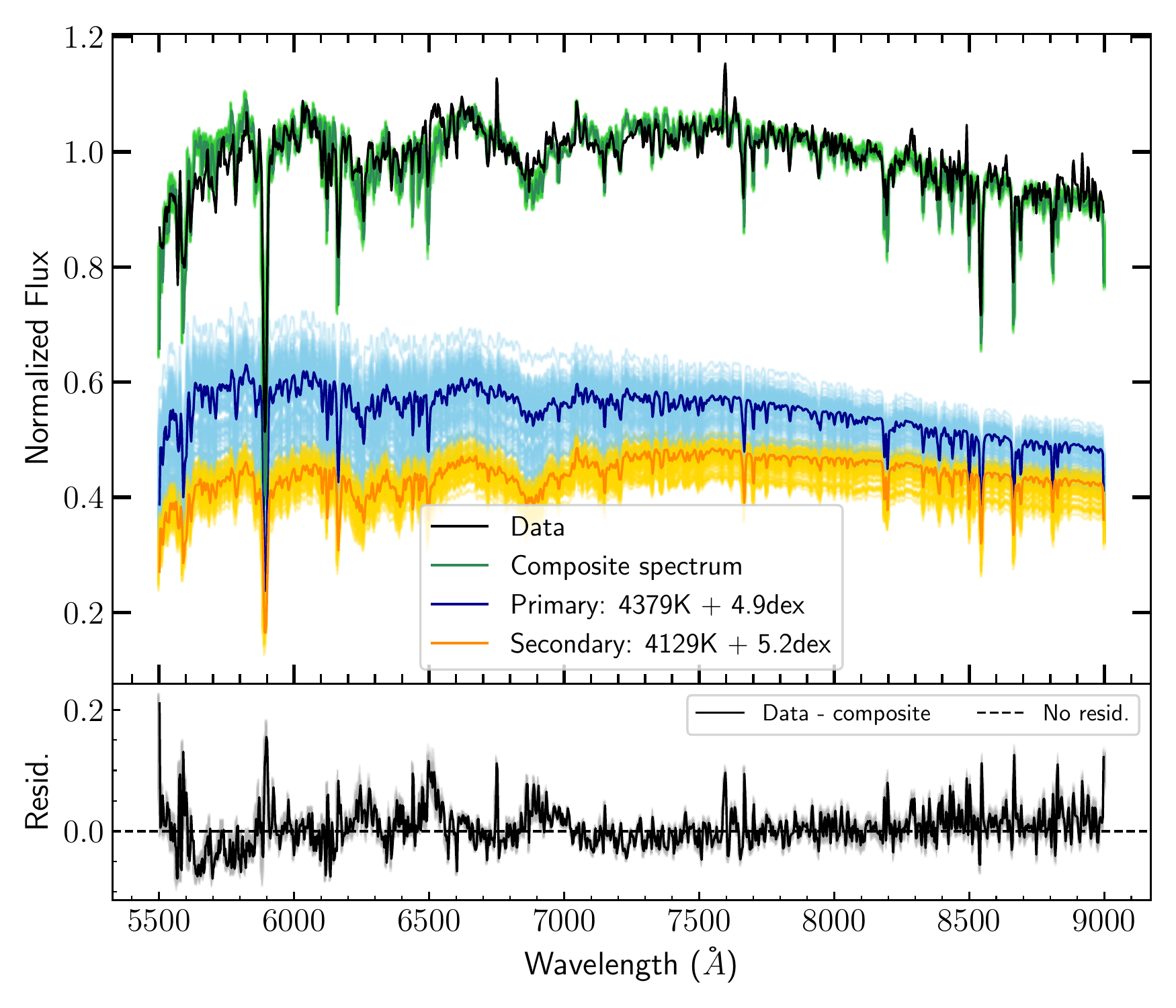} 
	\includegraphics[width = 0.45\linewidth]{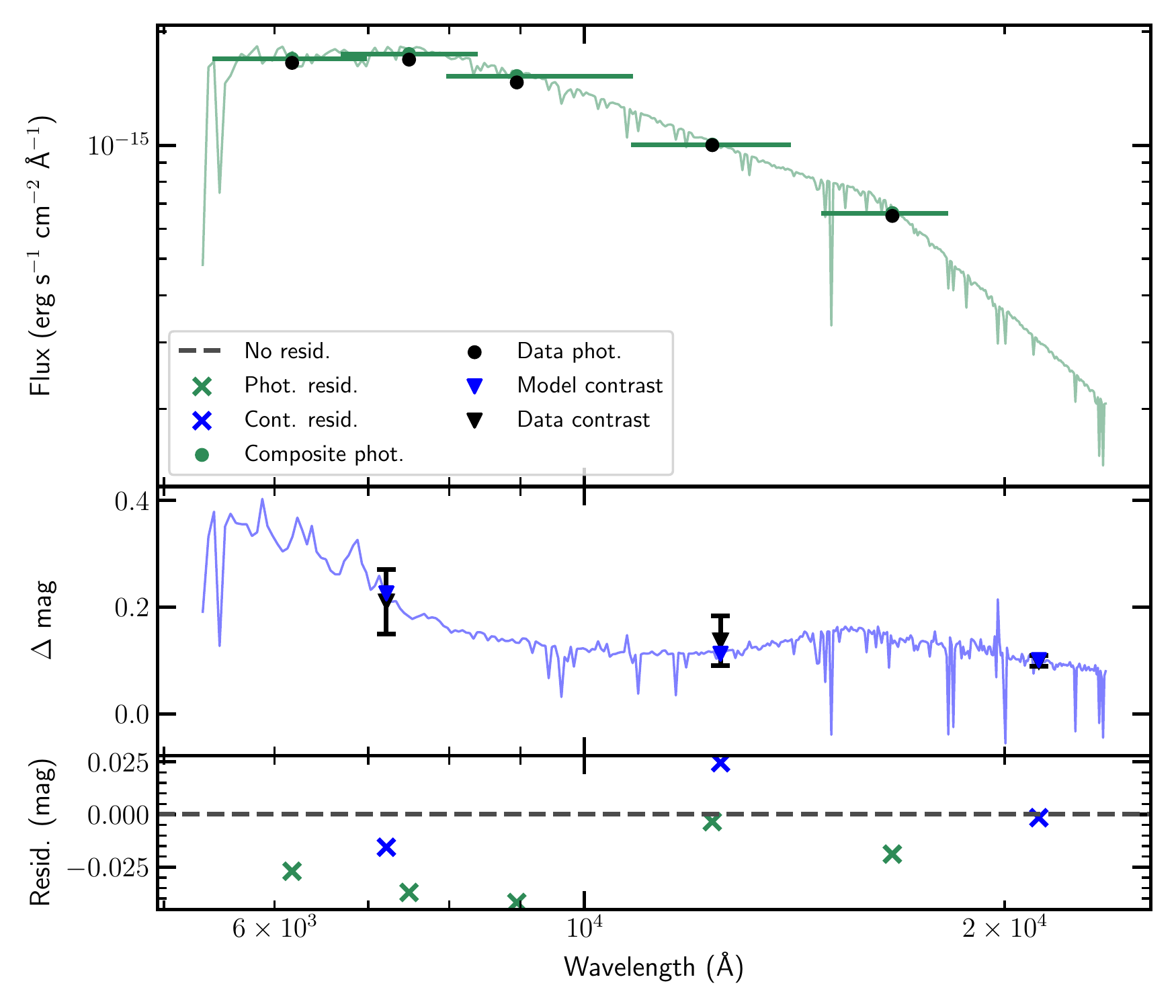}
}
\gridline{\includegraphics[width = 0.8\linewidth]{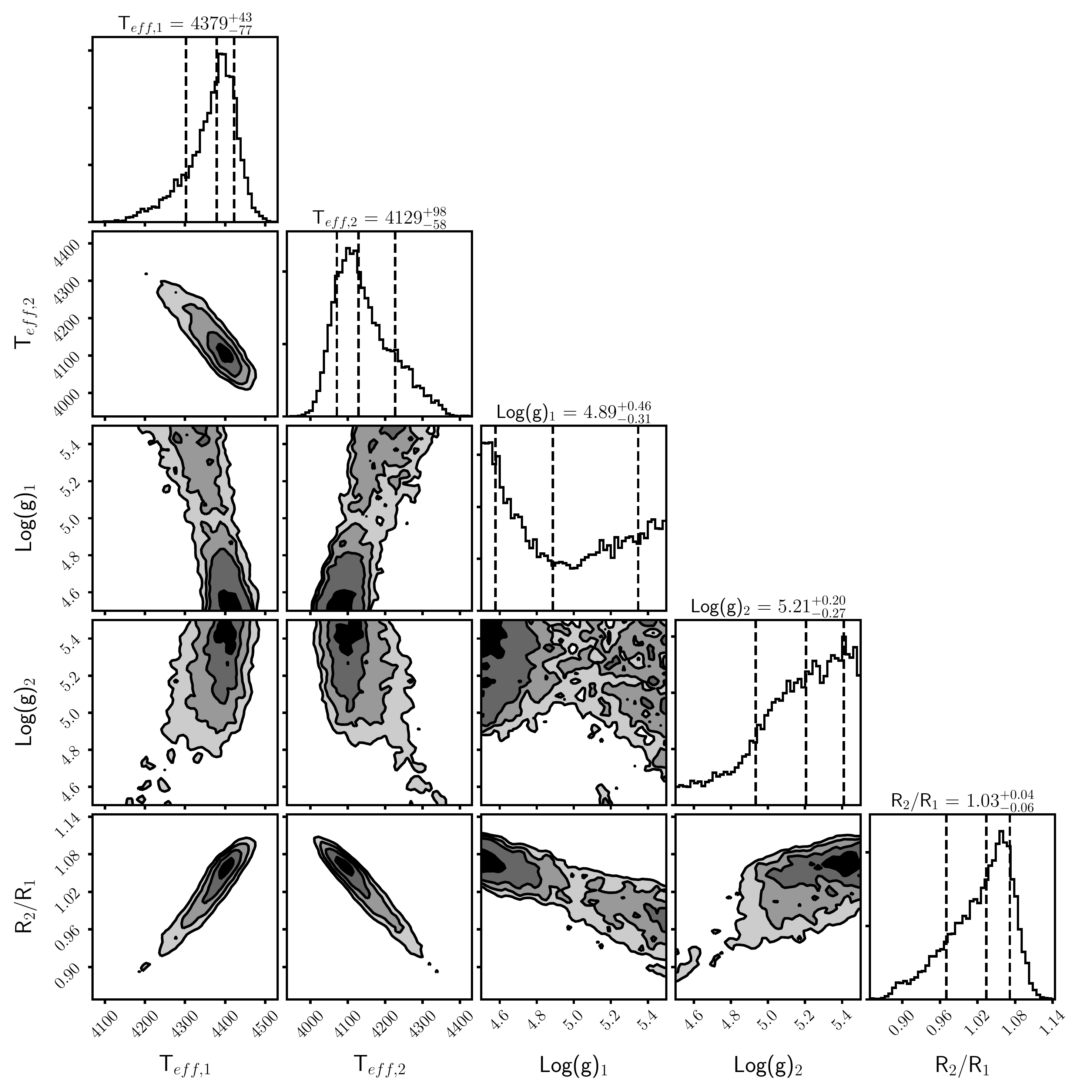}}
\caption{KOI 2174}
\label{fig:2174 spec}
\end{figure*}

\begin{figure*}
\gridline{\includegraphics[width = 0.45\linewidth]{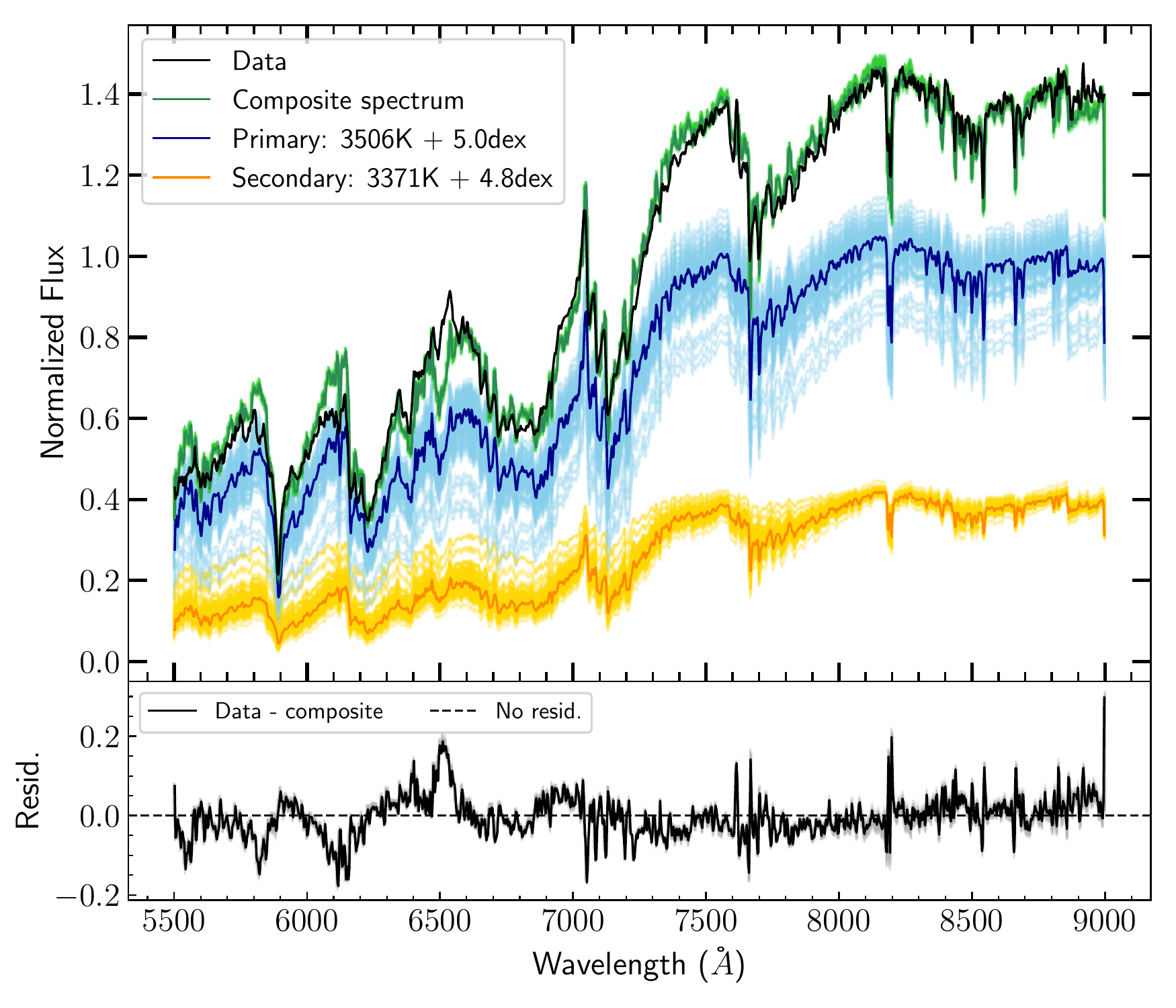}
	\includegraphics[width = 0.45\linewidth]{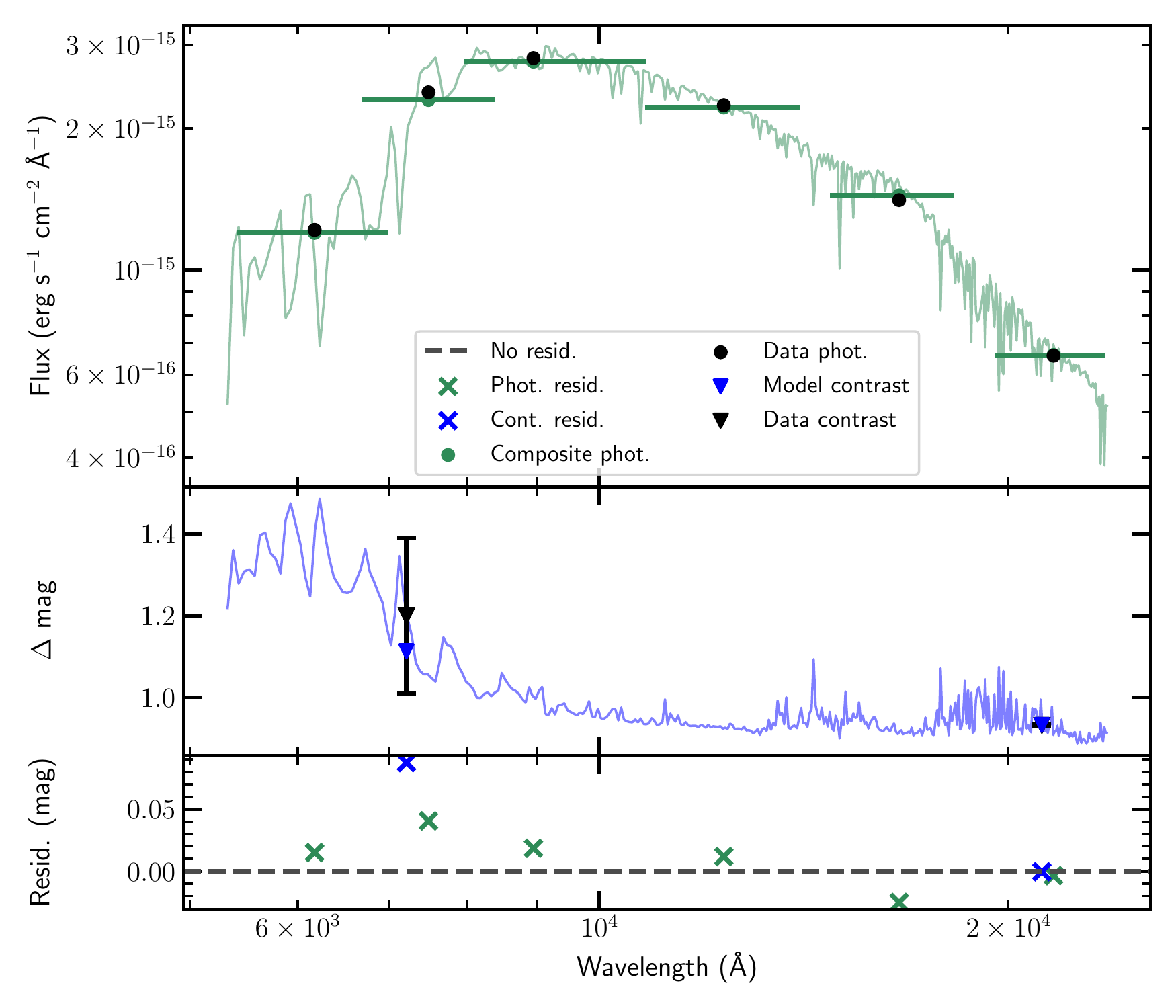}
}
\gridline{\includegraphics[width = 0.8\linewidth]{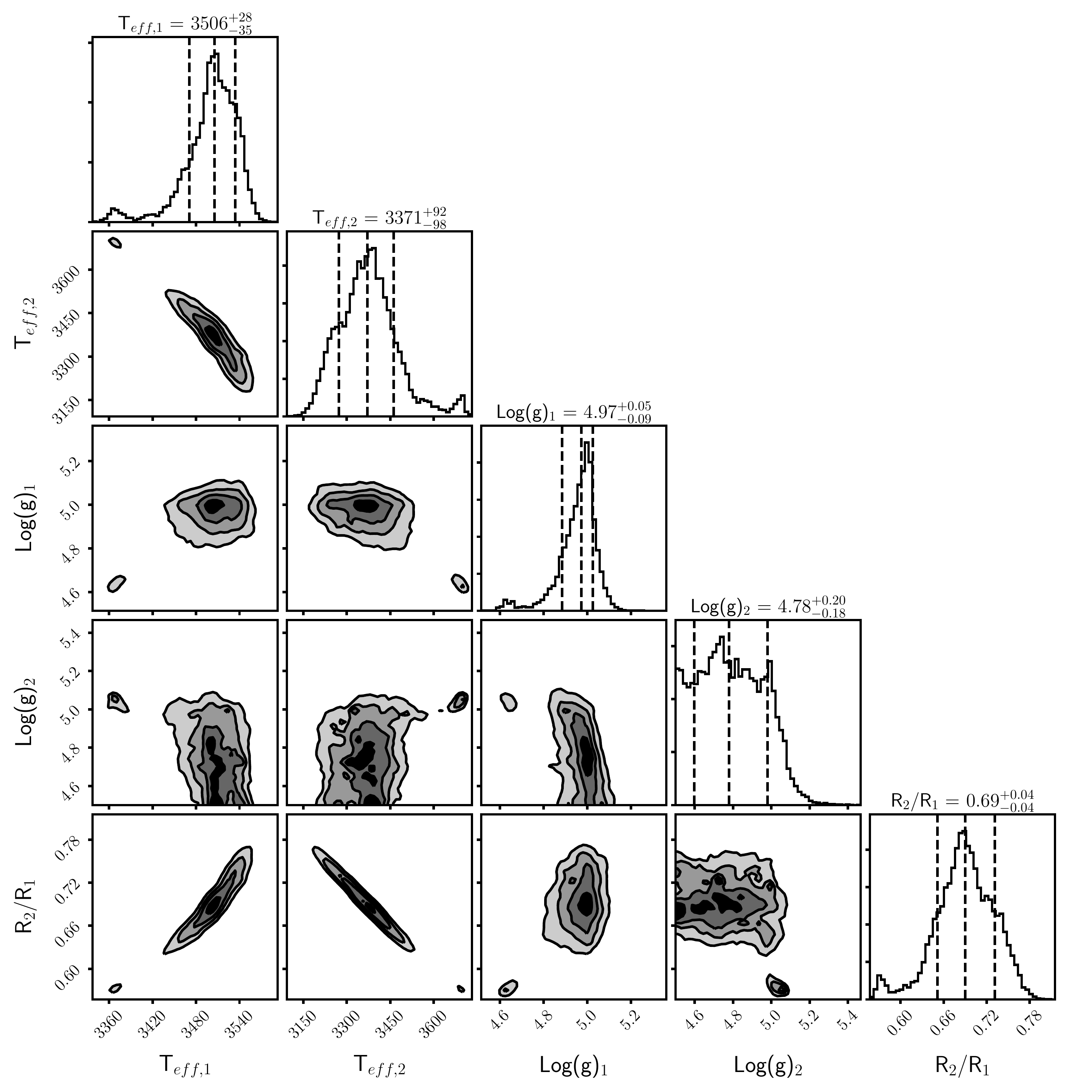}}
\caption{KOI 2542}
\label{fig:2542 spec}
\end{figure*}

\begin{figure*}
\gridline{\includegraphics[width = 0.45\linewidth]{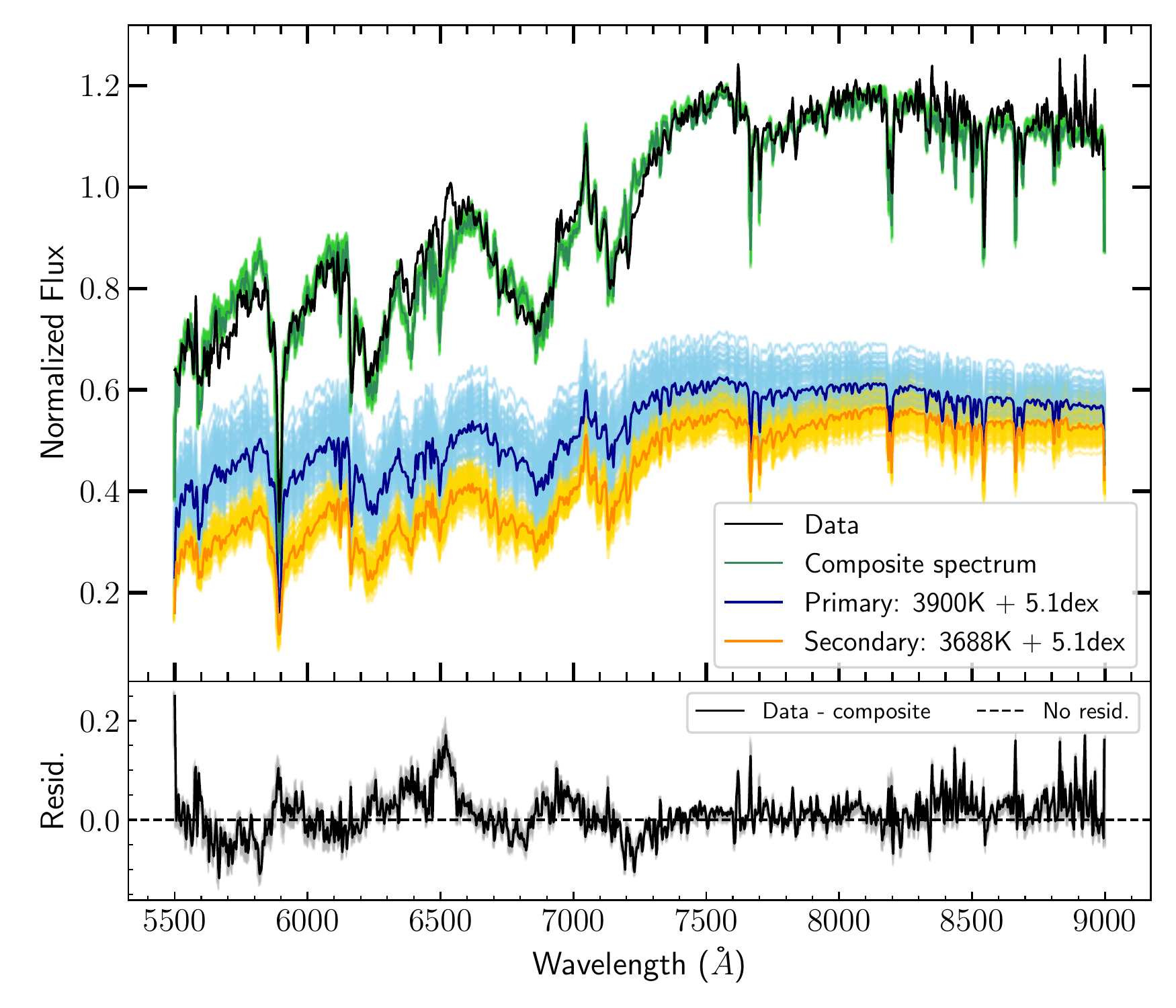} 
    \includegraphics[width = 0.45\linewidth]{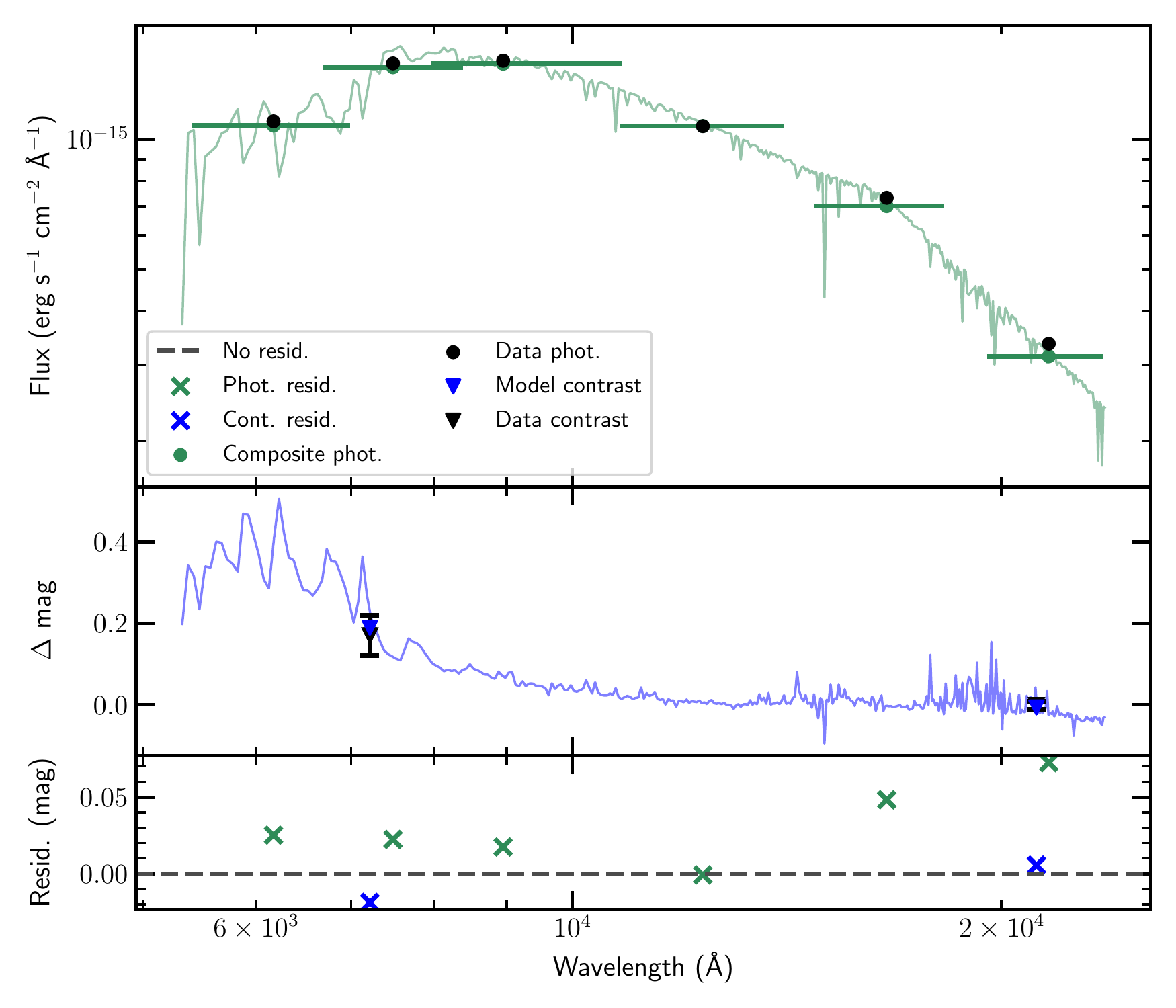}
}
\gridline{\includegraphics[width = 0.8\linewidth]{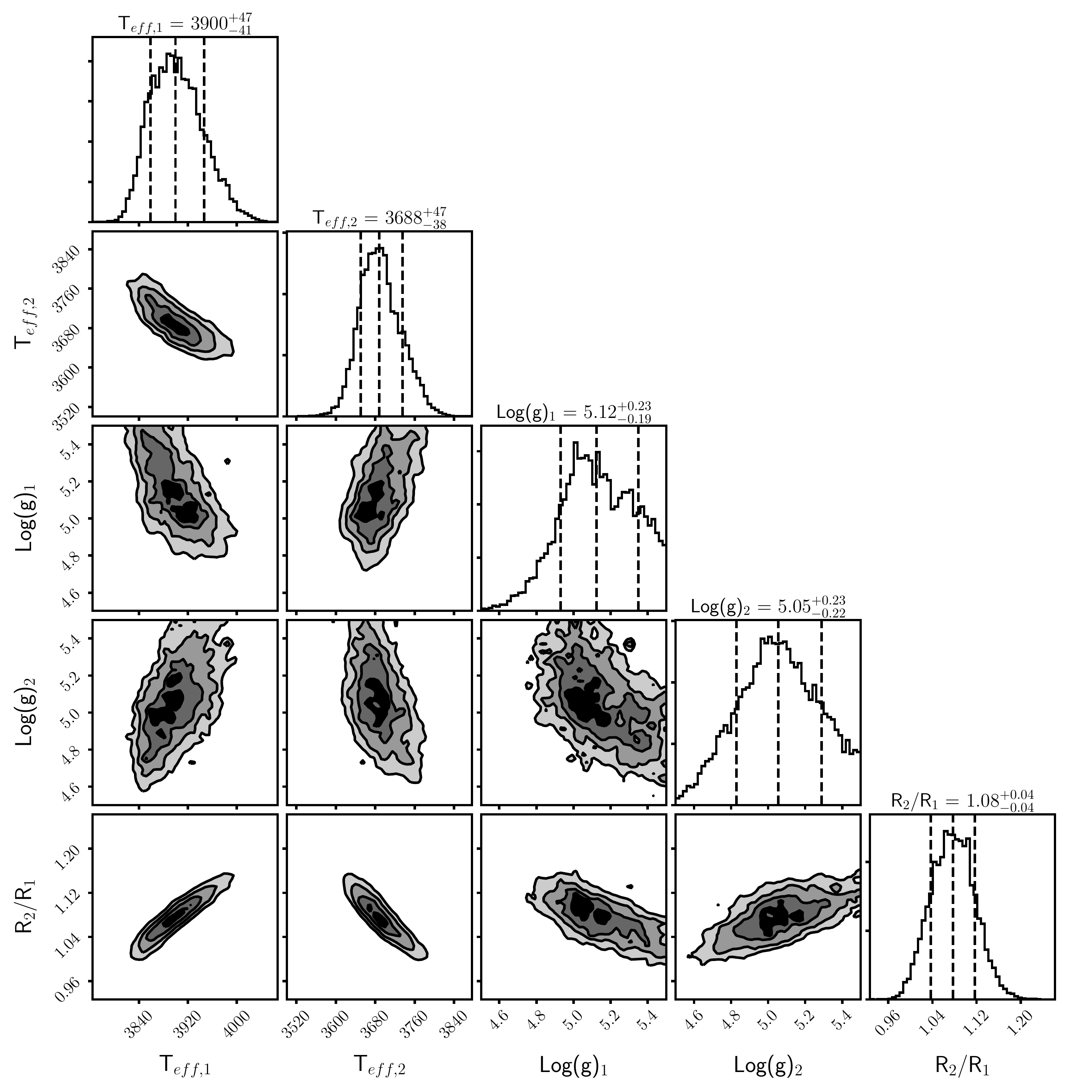}}
\caption{KOI 2862}
\label{fig:2862 spec}
\end{figure*}

\begin{figure*}
\gridline{\includegraphics[width = 0.45\linewidth]{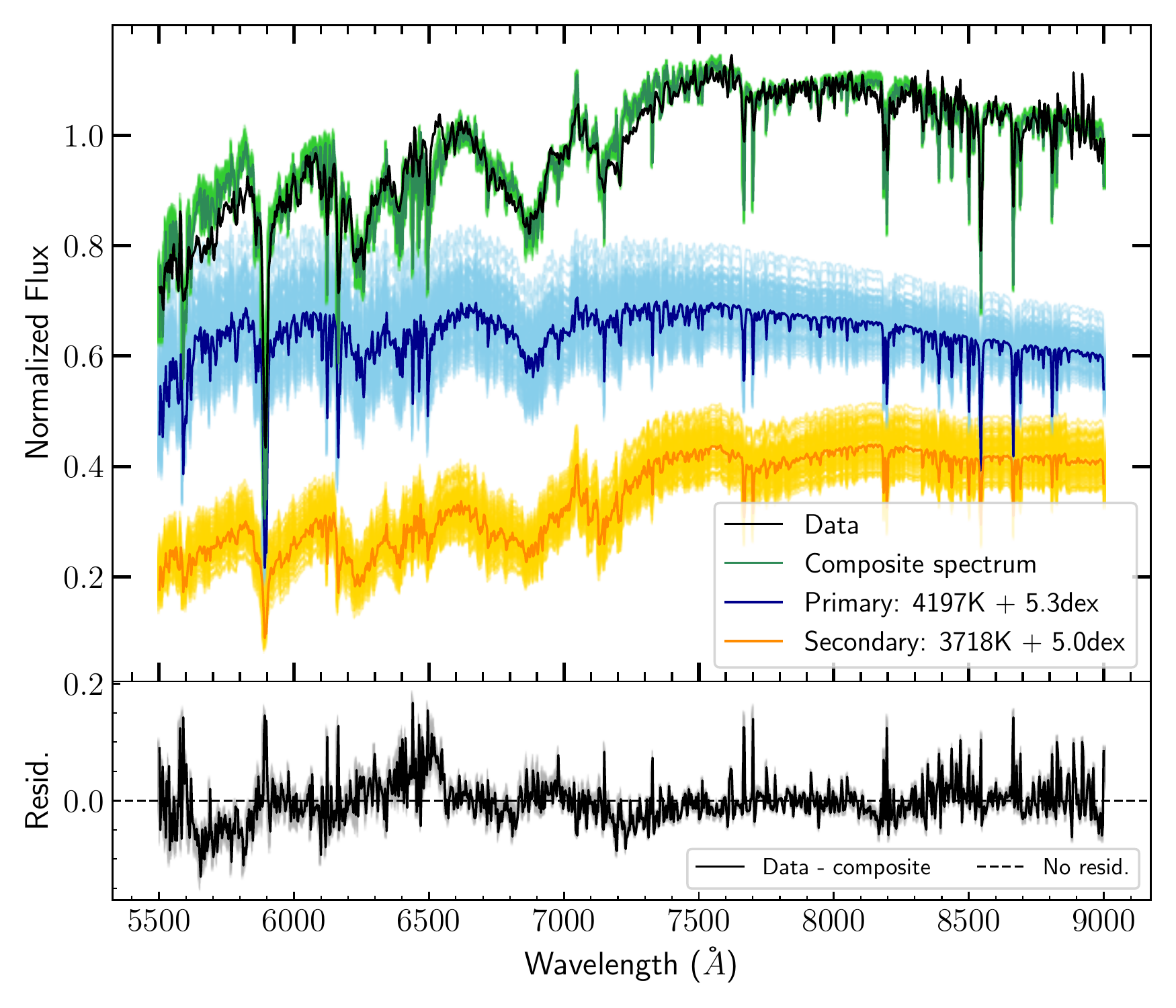} 
	\includegraphics[width = 0.45\linewidth]{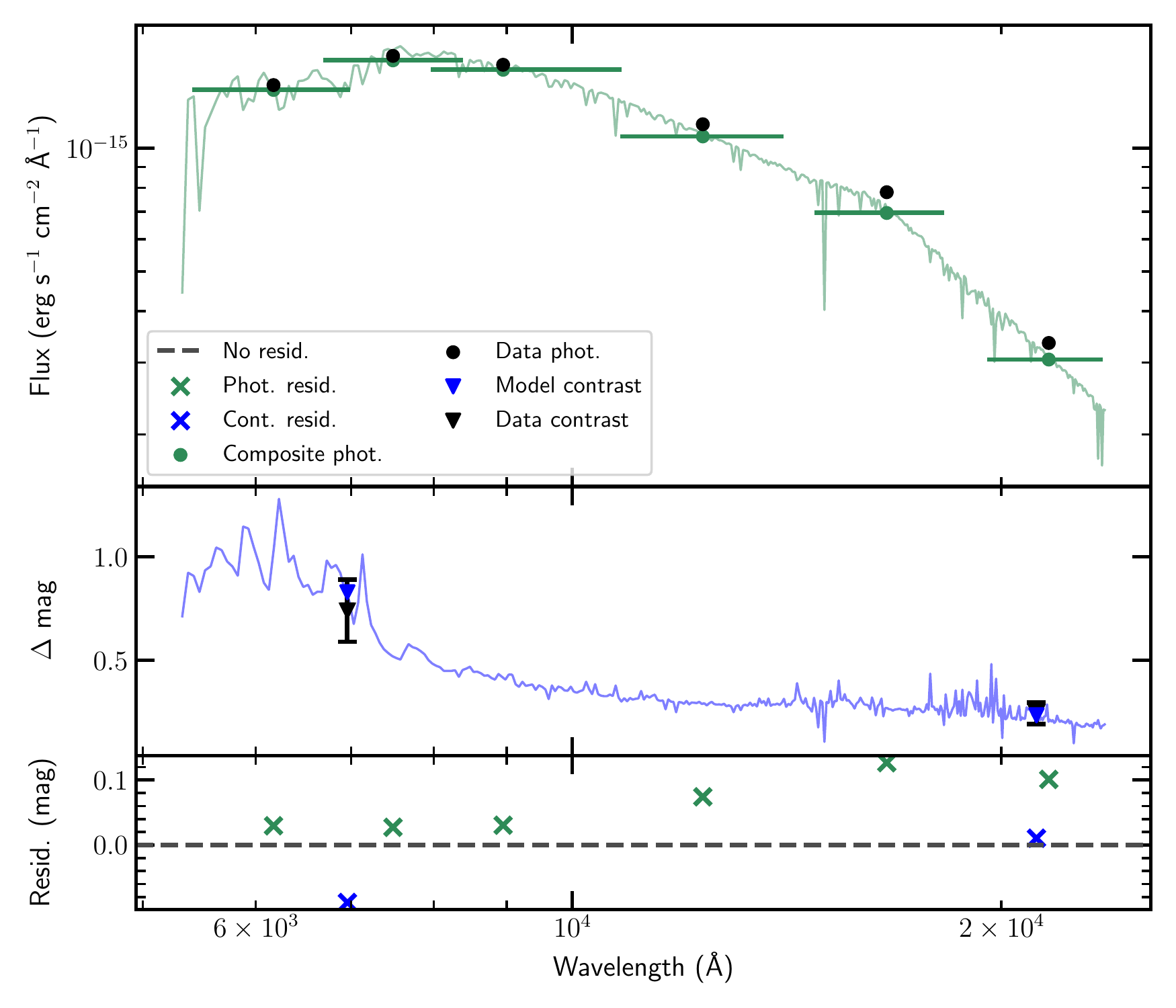}
}
\gridline{\includegraphics[width = 0.8\linewidth]{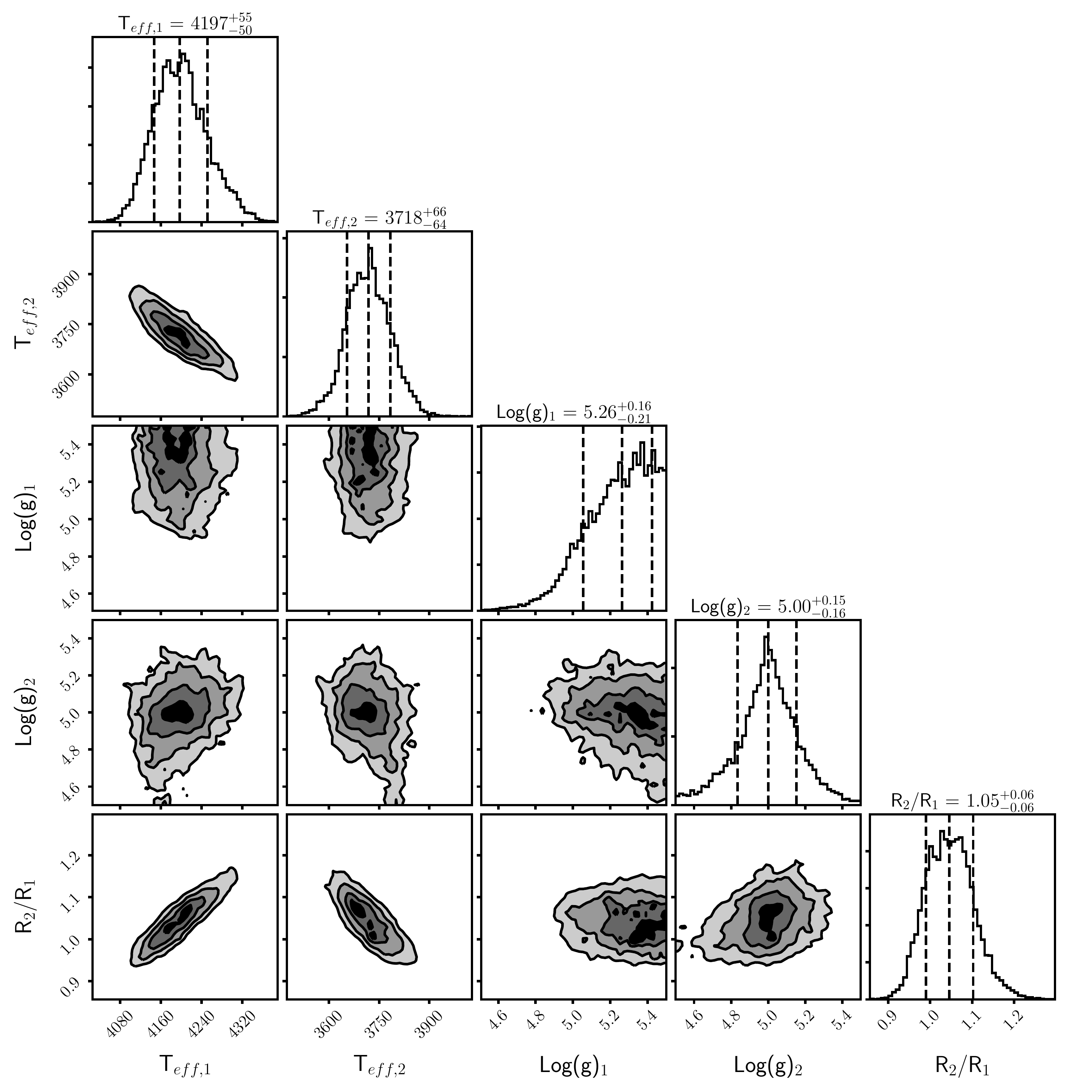}}
\caption{KOI 3010}
\label{fig:3010 spec}
\end{figure*}

\bibliography{mcmc_bib}
\end{document}